\documentclass[showpacs,amsfonts,aps,prc,nofootinbib,floatfix,superscriptaddress]{revtex4}

\usepackage{mathrsfs}
\usepackage{epsfig}
\usepackage{amssymb}
\usepackage{amsfonts}
\usepackage{latexsym}
\usepackage{amsthm}

\usepackage{bm}
\usepackage{graphicx}
\usepackage{amsmath}
\usepackage{hyperref}
\usepackage{slashed}

\def\p{{\cal P}}

\begin{document}

%%%%%%%%%%%%%%%%%%%%%%%%Front Matter%%%%%%%%%%%%%%%%%%%%
%%%%%%%%%%%%%%
%%%%%%%%%%%%%%%%%%%%%%%%%%%%%%%%%%%%%%%%%%%%%%%%%%
%%%%%%%%%%%%%%%%%%%%

\title{Photon Emission from a Momentum Anisotropic Quark-Gluon Plasma}

\author{Chun Shen}
\email[Corresponding author:\ ]{chunshen@physics.mcgill.ca}
\affiliation{Department of Physics, McGill University, 3600 University Street, Montreal, Quebec, H3A 2T8, Canada}
\affiliation{Department of Physics, The Ohio State University,
  Columbus, Ohio 43210-1117, USA}
\author{Jean-Fran\c{c}ois Paquet}
\affiliation{Department of Physics, McGill University, 3600 University Street, Montreal, Quebec, H3A 2T8, Canada}
\author{Ulrich Heinz}
\affiliation{Department of Physics, The Ohio State University,
  Columbus, Ohio 43210-1117, USA}
\author{Charles Gale}
\affiliation{Department of Physics, McGill University, 3600 University Street, Montreal, Quebec, H3A 2T8, Canada}
\affiliation{Frankfurt Institute for Advanced Studies, Ruth-Moufang-Str. 1, D-60438 Frankfurt am Main, Germany}
  
\begin{abstract}
We compute the photon emission rate from a quark-gluon plasma with an anisotropic particle momentum distribution induced by a non-vanishing local shear pressure tensor. Our calculation includes photon production through Compton scattering and quark-antiquark annihilation at leading order in $\alpha_s$, with all off-equilibrium corrections to leading order in the momentum anisotropy. For fermions we prove that the Kubo-Martin-Schwinger (KMS) relation holds in the hard loop regime for any particle momentum distribution function that is reflection-symmetric. This supports the equivalence, for 2 to 2 scattering processes, of the diagrammatic and kinetic approaches to calculating the photon emission rate. We compare the viscous rates from these two approaches at weak and realistic coupling strengths and provide parameterizations of the equilibrium and viscous photon emission rates for phenomenological studies in relativistic heavy-ion collisions.
\end{abstract}


\date{\today}

\maketitle

%%%%%%%%%%%%%%%%%%%%%%%%%%%%%%%%%%%%%%%%%%%%%
\section{Introduction}
\label{sec1}
%%%%%%%%%%%%%%%%%%%%%%%%%%%%%%%%%%%%%%%%%%%%%

Heavy-ion collisions at the Relativistic Heavy-Ion Collider (RHIC) and the Large Hadron Collider (LHC) offer a privileged window for studying the physics of hot and dense strongly interacting matter. The smallness of the electromagnetic coupling constant and small extent of the hot QCD medium produced in heavy-ion collisions makes the latter largely transparent to electromagnetic probes such as thermal photons and dileptons. This is to be contrasted with the very small mean free path of colored particles in the medium. This difference means that, through their production rates in the medium, electromagnetic probes can provide information about the entire space-time evolution of the QCD medium that is not subsequently scrambled by further interactions.

Controlled calculations of the rate of photon production from a hot QCD medium are possible only in certain limiting situations. For a perfectly thermalized, weakly-coupled ($g_s{\,\ll\,}1$) quark gluon plasma (QGP) a complete calculation of the rate at $\mathcal{O}(e^2 g_s^2)$ has been available for a decade \cite{Arnold:2001ms}. The next-to-leading-order correction $\mathcal{O}(e^2 g_s^3)$ to the thermal photon rate was computed recently \cite{Ghiglieri:2013gia}. At temperatures below the pseudocritical temperature for the quark-hadron phase transition, $T_\mathrm{c}{\,\sim\,}155{-}165$\,MeV, where dense QCD matter is modeled as a hadron resonance gas, effective Lagrangian approaches have been adopted \cite{Turbide:2003si}. Those calculations assume that the medium is static, homogeneous, and fully thermalized.

The success of hydrodynamical descriptions of the hot QCD medium created in heavy ion collisions \cite{Heinz:2013th,Gale:2013da} makes it reasonable to assume that the medium is not too far from local thermal equilibrium. However, non-zero values for its transport coefficients, resulting from non-zero mean free paths of the constituents, lead to deviations from local thermal equilibrium in an expanding system which increase with the expansion rate. For example, in an anisotropically expanding system shear viscosity causes the momentum distribution in the local rest frame to become anisotropic itself, falling off more steeply in the directions into which the system expands more rapidly.

A number of attempts have been made at evaluating the consequences of such off-equilibrium effects on the (virtual) photon emission rates in a QGP \cite{Dusling:2008xj,Dusling:2009bc,Dion:2011pp}. However, these previous works all share one shortcoming: for a given collision process that results in the emission of a photon, they include the viscous corrections to the local momentum distribution functions only for the incoming and outgoing particles, but ignore viscous medium modifications of the collision matrix element itself. For scattering processes in which the inclusion of medium effects is essential (for example, when dynamical mass generation for the medium constituents serves as a regulator for infrared divergences associated with otherwise massless particle exchange) viscous corrections to the distribution functions can lead to significant modifications of the screening mechanism and therefore to the collision matrix element. This problem was first tackled in \cite{Baier:1997xc,Schenke:2006fz,Schenke:2006yp} for simple parameterizations of the local momentum anisotropy. 

The present paper builds on these publications and considers a more general ansatz of momentum anisotropy, namely
\begin{equation}
f(K) \equiv f_0(k) +\delta f(k) = f_0(k)\left[1 + \bigl(1{\pm}f_0(k)\bigr) \frac{\pi^{\mu\nu} \hat{k}_\mu \hat{k}_\nu }{2(e{+}\p)} \chi(k/T)\right].
\label{eq1}
\end{equation}
Here $e$, $\p$, $T$, and $\pi^{\mu\nu}$ are functions of space-time position $x$ denoting the local energy density, pressure, temperature, and shear stress tensor of the expanding medium. The particle's energy in the local rest frame is $k = K \cdot u$, $\hat{k}^\mu \equiv K^\mu/k$ is a light-like vector with unit time component in the local rest frame, and the scalar function $\chi(k/T) = (k/T)^\lambda$ with $1{\,\le\,}\lambda{\,\le\,}2$ controls the energy dependence of the off-equilibrium correction. The form (\ref{eq1}) for the deviation $\delta f$ from local equilibrium, as well as the quoted range for the parameter $\lambda$, follows from a solution of the kinetic equation for $f(k)$ with a Boltzmann collision term that has been linearized around local equilibrium $f_0(k)$ \cite{Dusling:2009df}; the exponent $\lambda$ is related to the energy dependence of the differential scattering cross section. For a system whose non-equilibrium transport properties are dominated by shear viscosity, equation~(\ref{eq1}) is sufficiently general to describe the momentum distribution of particles in a weakly coupled expanding plasma as long as its $\pi^{\mu\nu}$ is not too large \cite{Baier:2006um}. Other transport effects, such as bulk viscosity and heat conductivity, are neglected in this work, and the baryon chemical potential is assumed to vanish. We note however that the methods used in this paper should be generalisable to different $\delta f$, such as that associated with bulk viscosity.

We will assume that the spatial dependence of the medium is sufficiently weak that all space-time gradient effects can be accounted for through the shear stress tensor $\pi^{\mu\nu}$. This means that $f(x,K)$ depends on $x$ only parametrically (through $T(x)$, $\pi^{\mu\nu}(x)$, etc.), and we will henceforth drop the $x$-dependence of $f$. Space-time integrals in the evaluation of Feynman diagrams occurring in the computation of the emission rate at point $x$ will be done as if the system were infinite and static (with the given values for $T$, $\pi^{\mu\nu}$ etc.), i.e. we will continue to assume that energy and momentum are conserved in any scattering process. This corresponds to the assumption that photon emission is local on length scales that characterize the space-time variability of the emitting medium. 

We consider here photon production from leading order (in $\alpha_s$) $2\to 2$ processes only.
We further linearize the viscous correction in the shear stress tensor, yielding a result accurate to leading order in $\pi^{\mu\nu}/(e+\p)$.
The inclusion and calculation of viscous corrections for the family of soft $2\to n$ diagrams that are required for an evaluation of the photon emission rate to full leading order in $\alpha_s$ \cite{Arnold:2001ms} is left for future work.

The paper is structured as follows: In Sec.~\ref{sec2} we present the calculation of the viscous corrections to the QGP photon emission rate. In Sec.~\ref{sec2a} we introduce a tensor decomposition technique to isolate the linear off-equilibrium correction coefficient and write the rate in the convenient form
\begin{equation}
\label{eq2}
 k \frac{dR}{d^3k} = T^2 \left( \tilde\Gamma_0 + 
                                               \frac{\pi^{\mu\nu} \hat{k}_\mu \hat{k}_\nu}{2(e{+}\p)}\,
                                               \tilde\Gamma_1\right).
\end{equation}
where both the thermal equilibrium rate $\tilde\Gamma_0{\,\equiv\,}\Gamma_0/T^2$ and the viscous coefficient $\tilde\Gamma_1{\,\equiv\,}\Gamma_1/T^2$ (see Eqs.~(\ref{eq11}), (\ref{eq12}) below) are dimensionless scalar functions of the normalized photon local rest frame energy $\kappa=k/T \equiv u{\cdot}K/T$. In Sec.~\ref{sec2b} we compute $\tilde\Gamma_0$ and $\tilde\Gamma_1$ using a diagrammatic approach, starting with a proof of the KMS relation for the fermionic self energy to leading order in a high-temperature (soft external momentum or hard thermal loop (HTL)) approximation. This KMS relation is necessary for the equivalence of the photon emission rate calculated in the diagrammatic approach with the kinetic theory calculation that we present in Sec.~\ref{sec2c}. In Sec.~\ref{sec3} we evaluate the rates and viscous correction coefficients numerically for both weak and realistically strong coupling $\alpha_s$. In particular, we explore the sensitivity of the diagrammatic calculation, which is split into a soft and hard exchanged momentum contribution, on the cutoff momentum separating the soft and hard regions. The kinetic approach effectively implements an alternate resummation scheme for subleading terms that are higher order in $g_s$ and does not require such a cutoff. Comparing the two approaches quantitatively, we use the difference between the corresponding equilibrium rates and viscous correction coefficients as a measure to gauge the theoretical uncertainty of our result. Conclusions and final comments are offered in Sec.~\ref{sec4}. Some technical details of the calculation in Sec.~\ref{sec2b} are relegated to Appendix~\ref{appHard}.

For convenience, a parametrization of $\tilde\Gamma_0$ and $\tilde\Gamma_1$ is given in Appendix~\ref{appendixFit}.

%%%%%%%%%%%%%%%%%%%%%%%%%%%%%%%%%%%%%%%%%%%%%
\section{Photon emission rates}
\label{sec2}
%%%%%%%%%%%%%%%%%%%%%%%%%%%%%%%%%%%%%%%%%%%%%

%%%%%%%%%%%%%%%%%%%%%%%%%%%%%%%%%%%%%%%%%%%%%
\subsection{General Formalism}
\label{sec2a}
%%%%%%%%%%%%%%%%%%%%%%%%%%%%%%%%%%%%%%%%%%%%%

The photon emission rate for a static medium is given in the real-time or Closed Time Path (CTP) formalism \cite{Rammer:2007zz} by \cite{Baier:1997xc,Serreau:2003wr}
\begin{equation}
   k \frac{d R}{d^3 k} = \frac{i}{2(2\pi)^3} (\Pi_{12}(K))^\mu\,_\mu.
\label{2.A.1}
\end{equation}
where $1$ ($2$) refers to the (anti-)time-ordered contour branch in the CTP formalism.
If the medium is in thermal equilibrium, the different components in the real-time formalism of the photon self-energy $\Pi^{\mu\nu}$ are related by the fluctuation-dissipation theorem. This is a consequence of the Kubo-Martin-Schwinger (KMS) condition satisfied by thermal equilibrium propagators due to their (anti-)periodicity in imaginary time with period $\beta=1/T$ \cite{LeBellac,Kapusta:2006pm}. Using this property of the photon self-energy, the photon emission rate can be written:
\begin{equation}
   k \frac{d R}{d^3 k} = - \frac{\mathrm{Im}\, (\Pi_\mathrm{ret}(K))^\mu\,_\mu}{(2\pi)^3(e^{K^0/T}{-}1)},
\label{2.A.2}
\end{equation}
where $\Pi_\mathrm{ret}$ is the retarded photon self-energy.

Finite temperature cutting rules for the calculation of the imaginary part of the retarded photon self-energy \cite{Weldon:1983jn,Gelis:1997zv,Majumder:2001iy,CaronHuotMSc} allow to rewrite the rate in kinetic theory form \cite{McLerran:1984ay}. For a process with $m$ incoming particles with four-momenta $P_1,\dots,P_m$ colliding to produce $n$ outgoing particles with momenta $P_{m{+}1},\dots,P_{m{+}n}$ plus a photon with momentum $K$, the contribution to the photon emission rate is   
\begin{eqnarray}
\label{2.A.3}
  k \frac{d R}{d^3 k} &=& N \int \frac{d^3p_1}{2E_1(2\pi)^3} \cdots \frac{d^3p_m}{2E_m(2\pi)^3} 
  \cdots \frac{d^3p_{m+n}}{2E_{m+n}(2\pi)^3} (2\pi)^4 \delta^{(4)}\!\!
  \left(\sum_{i=1}^m P_i^\mu - \!\!\sum_{j=m+1}^{m+n} P_j^\mu-K^\mu\right)  
  \notag \\
  &&\times \vert {\cal M} \vert^2 f_{B/F}(P_1) \cdots f_{B/F}(P_m) (1 \pm f_{B/F}(P_{m+1})) \cdots (1 \pm f_{B/F}(P_{m+n})),
\end{eqnarray}
where $f_{B(F)}(P)$ are Bose (Fermi) distribution functions for bosons (fermions), and $N$ is an overall degeneracy factor that depends on the specific production channel.

For an imperfectly thermalized, anisotropically expanding medium the particles' momentum distributions are no longer isotropic in the local rest frame. Considering shear viscous effects and writing the distribution function $f$ as in Eq.~(\ref{eq1}), the deviation $\delta f$ from locally isotropic equilibrium contributes to the photon emission rate (\ref{2.A.3}) generally both through the thermal weights for the incoming and outgoing particles and through the (medium-modified) production matrix element $\cal M$. Assuming that the inverse Reynolds number for the shear stress tensor $\pi^{\mu\nu}$ is small, $\mathrm{Re}_\pi^{-1}\equiv\sqrt{\pi^{\mu\nu}\pi_{\mu\nu}}/(e{+}\p){\,\ll\,}1$, such that viscous fluid dynamics is applicable and $\delta f \ll f_0$, we can expand the photon emission rates in powers of $\pi^{\mu\nu}$:
\begin{equation}
\label{2.A.4}
   k \frac{d R}{d^3 k} = \Gamma_0 + \frac{\pi^{\mu\nu}}{2(e+\p)} \Gamma_{\mu\nu} + 
   {\cal O}\left(\left(\frac{\pi^{\mu\nu}}{2(e+\p)}\right)^2\right).
\end{equation}
Here $\Gamma_0$ stands for the thermal equilibrium emission rate while $\Gamma_{\mu\nu}$ is the rate coefficient of the first order viscous correction. Both $\Gamma_0$ and $\Gamma_{\mu\nu}$ involve only integrals over equilibrium distribution functions and, for a medium consisting of massless particles, are proportional to $T^2$ times dimensionless functions of the local rest frame photon energy in units of temperature, $k/T$. 

By definition $\pi^{\mu\nu}$ is symmetric, traceless and has only spatial components in the local rest frame. This is formally expressed in the identity 
\begin{equation}
\label{2.A.5}
  \pi^{\mu\nu}=\Delta^{\mu\nu}_{\alpha\beta}\pi^{\alpha\beta},
\end{equation}
with the symmetric, locally spatial and traceless projector
\begin{equation}
\label{2.A.6}
  \Delta^{\mu\nu}_{\alpha\beta} = \textstyle{\frac{1}{2}}\left(\Delta^\mu_\alpha
  \Delta^\nu_\beta{+}\Delta^\nu_\alpha\Delta^\mu_\beta\right) - \textstyle{\frac{1}{3}}\Delta^{\mu\nu}
  \Delta_{\alpha\beta},
\end{equation}
where $\Delta^{\mu\nu}=g^{\mu\nu}{-}u^\mu u^\nu$ and $g^{\mu\nu}=(1,-1,-1,-1)$.

The most general tensor decomposition of $\Gamma^{\mu\nu}$ involves symmetrized terms proportional to $g^{\mu\nu},\,u^\mu u^\nu,\, u^\mu \hat{k}^\nu$, and $\hat{k}^\mu \hat{k}^\nu$. Due to the properties of $\Delta^{\mu\nu}_{\alpha\beta}$, only the last term survives in the product $\pi^{\mu\nu}\Gamma_{\mu\nu}$ \cite{Dusling:2008xj}:
\begin{equation}
\label{eq11}
 \pi^{\mu\nu} \Gamma_{\mu\nu} = \pi^{\mu\nu}  \Delta_{\mu\nu}^{\alpha\beta}\Gamma_{\alpha\beta} =
 \Gamma_1 \pi^{\mu\nu} \hat{k}_\mu \hat{k}_\nu.
\end{equation}
The scalar coefficient $\Gamma_1$ can be obtained from $\Gamma^{\mu\nu}$ by contracting with
\begin{eqnarray}
   a_{\mu\nu} = \frac{1}{2} \Delta_{\mu\alpha}\Delta_{\nu\beta}
                         \left(g^{\alpha\beta} + 3\hat{k}^\alpha \hat{k}^\beta \right). 
\label{eq12}
\end{eqnarray}
This leads to Eq.~(\ref{eq2}), with $\tilde\Gamma_1{\,\equiv\,}a_{\alpha \beta} \Gamma^{\alpha \beta}/T^2$. 

It is worth noting that the structure (\ref{eq2}) of the photon emission rate is independent of the collision kernel and holds, to linear order in $\pi^{\mu\nu}$, for any medium. Medium properties enter only in the explicit calculation of the scalar functions $\Gamma_0$ and $\Gamma_1$.

Note that the factorization of the viscous correction to the emission rate (\ref{eq2}) into two Lorentz scalars, $\pi^{\mu\nu} \hat{k}_\mu \hat{k}_\nu$ and $\Gamma_1{\,=\,}a_{\alpha\beta}\Gamma^{\alpha\beta}$, is numerically advantageous since each of those scalars can be evaluated in a different reference frame. The term $\pi^{\mu\nu} \hat{k}_\mu \hat{k}_\nu$ can be computed in the laboratory frame used for solving the hydrodynamic equations where $\pi^{\mu\nu}(x)$ and the measured photon momentum are known directly. The factor $\Gamma_1$, on the other hand, is most easily worked out in the local rest frame where the thermal equilibrium distributions simplify. Numerically expensive repeated Lorentz boosts of vectors and tensors between frames are thus avoided. For phenomenological studies, the local equilibrium rate $\Gamma_0$ and viscous correction coefficient $\Gamma_1$ can both be tabulated or parametrized as functions of the local rest frame photon energy which is easily computed as $k{\,=\,}u{\cdot}K$ from the hydrodynamic flow velocity and measured photon momentum in the laboratory frame. We provide parameterizations of $\tilde{\Gamma}_0$ and $\tilde{\Gamma}_1$ in Appendix~\ref{appendixFit}.

%%%%%%%%%%%%%%%%%%%%%%%%%%%%%%%%%%%%%%%%%%%%%
\subsection{Diagrammatic approach}
\label{sec2b}
%%%%%%%%%%%%%%%%%%%%%%%%%%%%%%%%%%%%%%%%%%%%%

In the diagrammatic approach the calculation of the photon production rate starts from Eq.~(\ref{2.A.1}). The evaluation of $(\Pi_{12}(K))^\mu\,_\mu$ involves a momentum loop integral  split into two domains, referred to as the soft and hard parts, which are separated by a cutoff momentum $q_\mathrm{cut}$. 
We begin with the calculation of the soft part, which requires the use of HTL resummed quark propagators to properly take into account the effect of the thermal medium on particle exchanges.
In equilibrium, the use of the KMS relation simplifies this task significantly. 
(For the fermion propagator this relation reads \cite{Rammer:2007zz,LeBellac,Greiner:1998vd,Greiner:1998ri}
{$G_{12}(Q){\,=\,}{-}e^{-Q^0/T}G_{21}(Q)$.)
We therefore first demonstrate that the KMS relation continues to hold, to leading order in $g_s$, if thermal equilibrium distributions are replaced by the viscously deformed distribution (\ref{eq1}). The proof does not rely on $\pi^{\mu\nu}$ being small, but only on the momentum-reflection symmetry of Eq.~(\ref{eq1}) in the local rest frame, and on the smallness of $g_s{\,\ll\,}1$ such that one can replace $e^{-Q^0/T}{\,\approx\,}1$ for soft momentum $Q^0{\,\sim\,}{\cal O}(g_s T)$.  

%%%%%%%%%%%%%%%%%%%%%%%%%%%%%%%%%%%%%%%%%%%%%
\subsubsection{KMS-like relation for the fermionic self-energy with anisotropic momentum distributions}
\label{sec2b1}
%%%%%%%%%%%%%%%%%%%%%%%%%%%%%%%%%%%%%%%%%%%%%

Following \cite{Schenke:2006fz}, we calculate the one-loop off-diagonal quark self-energies $\Sigma^{12}(P)$ (see Fig.~\ref{KMS.fig1}) and $\Sigma^{21}(P)$ and check that the approximate KMS relation $\Sigma^{12}(P){\,=\,}{-}\Sigma^{21}(P)$ holds in the hard loop limit $K^0,|\bm{K}|{\,\sim\,}T{\,\gg\,}P^0{\,\sim\,}{\cal O}(g_s T)$.
%
%===================== Fig.1 ==================
\begin{figure}[h!]
  \centering
  \includegraphics[width=0.4\linewidth]{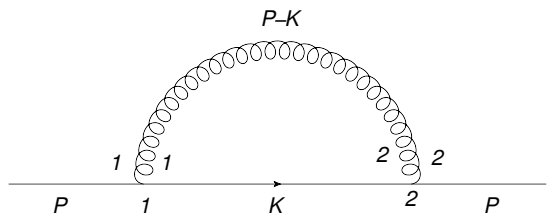}
  \caption{Off-diagonal component $\Sigma^{12}(P)$ of the one-loop quark self energy.}
  \label{KMS.fig1}
\end{figure}
%============================================
%
We write 
\begin{eqnarray}
\nonumber
- i \Sigma^{12}(P) &=& \int \frac{d^4K}{(2\pi)^4} (t^a t^a) (ig\gamma^\mu) iS_{12}(K)(-ig\gamma^\nu) (-i g_{\mu\nu} \Delta_{12}(P{-}K))
\end{eqnarray}
as
\begin{eqnarray}
  \Sigma^{12}(P) &=& - 2ig^2C_F \int \frac{d^4K}{(2\pi)^4} S_{12}(K) \Delta_{12}(P{-}K),
\label{KMS.1}
\end{eqnarray}
where $t^a$ are the $SU(3)$ gauge group matrices in the fundamental representation, and $S(K)$ and $\Delta(K)$ are the free fermion and scalar propagators,

\begin{eqnarray}
&&S(K) = \slashed{K} \left[ \left( \begin{array}{cc}
     \frac{1}{K^2 + i\epsilon} & 0 \\ 0 &  \frac{-1}{K^2 - i\epsilon} \end{array}\right) \right. \notag \\  
&&  \quad \quad \quad \quad \quad \quad +\left. 2\pi i \delta(K^2) \left( \begin{array}{cc}
     f_F(K) & -\theta(-K^0){+}f_F(K) \\ -\theta(K^0){+}f_F(K) &  f_F(K) \end{array}\right) \right]\!,  \quad
\label{KMS.2} \\
&&\Delta(K) = \left[ \left( \begin{array}{cc}
     \frac{1}{K^2 + i\epsilon} & 0 \\ 0 &  \frac{-1}{K^2 - i\epsilon} \end{array}\right) \right. \notag \\
&& \quad \quad \quad \quad \quad \quad \left. - 2\pi i \delta(K^2) \left( \begin{array}{cc}
     f_B(K) & \theta(-K^0){+}f_B(K) \\ \theta(K^0){+}f_B(K) &  f_B(K) \end{array}\right) \right].
\label{KMS.3}
\end{eqnarray}
Inserting the propagators into Eq.~(\ref{KMS.1}) we find
\begin{eqnarray}
  \Sigma^\mu_{12} (P) &=&- 2 i g_s^2 C_F \int \frac{d^4 K}{(2\pi)^4} K^\mu 2 \pi i \delta(K^2) 
  [-\theta(- K^0)+f_F(K)] 
  \notag \\
&&\times (-2 \pi i) \delta((P-K)^2) [\theta(- (P^0-K^0))+f_B(P-K)] \notag \\
&\approx& - 2 i g_s^2 C_F \int \frac{d^4 K}{(2\pi)^2} K^\mu \delta(K^2) [-\theta(- K^0)+f_F(K)] \delta(2 P \cdot K) [\theta(K^0)+f_B(K)].\quad
\label{KMS.4}
\end{eqnarray}
In the last step we assumed $P{\,\sim\,}{\cal O}(g_s T){\,\ll\,}K{\,\sim\,}{\cal O}(T)$. 
Letting $K^\mu \to - K^\mu$ and assuming that the off-equilibrium particle distribution functions satisfy $f_{B/F}(-K) = f_{B/F}(K)$, we find
\begin{eqnarray}
\Sigma^\mu_{12}(P) &=& 2 i g_s^2 C_F \int \frac{d^4 K}{(2\pi)^2} K^\mu \delta(K^2) [ - \theta(K^0) + f_F(K)]\delta(2 P \cdot K) [\theta(-K^0) + f_B(K)] 
\notag \\
\label{eq14}
&=& - \Sigma^\mu_{21}(P).
\label{eq:SigmaKMS}
\end{eqnarray}
This proves the desired relation. Note that our ansatz for the anisotropic momentum distribution, Eq.~(\ref{eq1}), respects the symmetry $f_{B/F}(-K) = f_{B/F}(K)$. This is easier to see in the fluid rest frame. By definition  $f_0(k)=1/(e^{-|K^0|/T}\pm 1)$, which is symmetric under $K \to -K$. The contraction $\pi^{\mu\nu} \hat{k}_\mu \hat{k}_\nu $ is also symmetric under reflection of $K$. Finally, $\chi(k/T)$ can be written as $(|K^0|/T)^\lambda$, $1<\lambda<2$, which again respects the necessary invariance.

We now proceed further to show that Eq.~(\ref{eq:SigmaKMS}) implies the validity of the fluctuation-dissipation theorem, which can be written as \cite{Rammer:2007zz,Greiner:1998vd,Greiner:1998ri} 
\begin{equation}
G_{12}(P)= \frac{2i}{e^{P^0/T}{+}1}\mathrm{Im}\,G_R(P) 
\end{equation}
where $G_R$ is the retarded fermion propagator and the prefactor reduces to a simple factor $i$ in the hard loop limit $P^0{\,\sim\,}{\cal O}(g_s T){\,\ll\,}T$. We start with the Dyson equation 
\begin{equation}
  G = G_0 + G_0 \Sigma G,
\label{eq15}
\end{equation}
where in the CTP formalism both the propagators and self-energy are $2 \times 2$ matrices. The (12)-component of the resummed propagator can be written in terms of $\Sigma_{12}$ and the retarded and advanced propagators $G_{R,A}$ as \cite{Rammer:2007zz,Greiner:1998vd,Greiner:1998ri} 
\begin{eqnarray}
  G_{12}(P) &=& G_R(P) \Sigma_{12}(P) G_A(P)  \notag \\
  &=& (-2 i) \frac{\Sigma_{12}(P)}{\Sigma_{21}(P) - \Sigma_{12}(P)} \mathrm{Im}\, G_R(P)
  = \frac{2i}{e^{P^0/T}{+}1} \mathrm{Im}\, G_R(P).
\end{eqnarray}
This is (a variant of) the fluctuation-dissipation theorem. With the KMS-like relation (\ref{eq14}) we see that in the hard loop limit it reduces to the simple form
\begin{equation}
 G_{12}(P) = i \mathrm{Im}\, G_R(P).
\end{equation}
We note that the validity of the KMS-like relation Eq. (\ref{eq14}) for anisotropic momentum distributions offers the possibility of generalizing results that were thought to be valid only in thermal equilibrium. In particular sum rule techniques developed in \cite{Ghiglieri:2013gia,CaronHuot:2008ni} will be useful to push to the next order in $g_s$ the photon rate production presented here.

%%%%%%%%%%%%%%%%%%%%%%%%%%%%%%%%%%%%%%%%%%%%%
\subsubsection{Retarded quark self-energy in near thermal equilibrium}
\label{sec2b2}
%%%%%%%%%%%%%%%%%%%%%%%%%%%%%%%%%%%%%%%%%%%%%

Equation~(\ref{eq14}) greatly simplifies the following calculations, by enabling us to relate the (12)-component of the photon self-energy $\Pi_{12}^{\mu\nu}$ to only the retarded quark self-energy
$\Sigma_R$.  

In the hard loop approximation, the retarded quark self-energy can be written as \cite{Mrowczynski:2000ed}
\begin{equation}
   \Sigma_R(P) = \frac{C_F}{4} g^2 \int \frac{d^3k}{(2\pi)^3} \frac{f(K)}{\vert \vec{k}\vert } 
   \frac{K \cdot \gamma}{K \cdot P + i \epsilon}
   \equiv \gamma_\mu\Sigma^\mu_R(P),
\label{selfenergy.1}
\end{equation}
where the $\gamma^\mu$ are the Dirac matrices and
\begin{equation}
   f(K) = 2(f_F(K) + \bar{f}_F(K)) + 4 f_B(K).
\label{selfenergy.2}
\end{equation}
In chemical equilibrium at zero net baryon density, $f_F(K) = \bar{f}_F(K)$. Hence,
\begin{equation}
  \Sigma_R(P) = C_F g^2 \int \frac{k dk}{2 \pi^2} \frac{d \Omega_k}{4 \pi} 
  (f_F(K) + f_B(K)) \frac{\hat{k} \cdot \gamma}{ \hat{k} \cdot P + i \epsilon},
\label{selfenergy.3}
\end{equation}
remembering that $\hat{k}^\mu = K^\mu/k$. 
The evaluation of $\Sigma_R(P)$ in the momentum-isotropic case is done e.g.~in~\cite{Blaizot:2001nr}.
Here we insert the anisotropic distribution function as in Eq. (\ref{eq1}):
\begin{eqnarray}
  \Sigma^\mu_R (p^0, p) &=& 
  \frac{C_F g^2}{2 \pi^2} \int kdk (f_{F0}(k) + f_{B0}(k)) \int \frac{d \Omega_k}{4 \pi} 
  \frac{\hat{k}^\mu}{\hat{k} \cdot P + i \epsilon} 
\notag \\
   &+& \frac{C_F g^2}{2 \pi^2} \frac{\pi_{\alpha \beta}}{2(e + \p)} \int k dk \,
   \bigl[f_{F0}(k)(1{-}f_{F0}(k)) + f_{B0}(k)(1{+}f_{B0}(k))\bigr]\, 
   \chi\Bigl(\frac{k}{T} \Bigr) 
\notag \\
   &\times& \int \frac{d \Omega_k}{4 \pi} 
   \frac{\hat{k}^\alpha \hat{k}^\beta \hat{k}^\mu}{\hat{k} \cdot P + i \epsilon}.
\label{selfenergy.4}
\end{eqnarray}
Note that the additional term is linear in $\pi_{\alpha \beta }$. We write 
\begin{equation}
   \Sigma_R^\mu(Q) = \Sigma^\mu_0(Q) + \frac{\pi_{\alpha\beta}}{2(e+\p)} 
   \Sigma_1^{\alpha \beta \mu}(Q) 
\label{soft.8}
\end{equation}
with
\begin{eqnarray}
   \Sigma_1^{\alpha \beta \mu} \equiv \frac{C_F g^2}{2 \pi^2} \int k dk \,
   \bigl[f_{F0}(k)(1{-}f_{F0}(k)) + f_{B0}(k)(1{+}f_{B0}(k))\bigr] \,\chi\Bigl(\frac{k}{T} \Bigr) 
%\notag \\ &\times&
   \int \frac{d \Omega_k}{4 \pi} 
   \frac{\hat{k}^\alpha \hat{k}^\beta \hat{k}^\mu}{\hat{k} \cdot P + i \epsilon}.\quad
  % \notag \\
\end{eqnarray}
For a given choice of $\chi\left(\frac{k}{T}\right)$, the $k$ integral can be evaluated and yields a pure number that we denote as $C_\mathrm{neq}$:
\begin{equation}
   \Sigma_1^{\alpha \beta \mu}(P) = \frac{C_F g^2 T^2}{2 \pi^2} C_\mathrm{neq} 
   \int \frac{d \Omega_k}{4 \pi} \frac{\hat{k}^\alpha \hat{k}^\beta \hat{k}^\mu}{\hat{k} \cdot P + i \epsilon}.
\end{equation}
Using tensor decomposition and the tracelessness and transversality of $\pi_{\alpha\beta}$ again, we write
\begin{equation}
   \pi_{\alpha \beta} \Sigma_1^{\alpha \beta \mu}(P) = \pi_{\alpha \beta} \left[ A_1(P) 
   \hat{p}^\alpha \hat{p}^\beta \hat{p}^\mu + B_1(P) \hat{p}^\alpha \hat{p}^\beta u^\mu 
   + C_1(P) (\hat{p}^\alpha g^{\beta\mu} + \hat{p}^\beta g^{\alpha\mu}) \right],
\end{equation}
where $\hat{p}^\mu{\,=\,}P^\mu/p$ with $p=|\bm{P}|$. The coefficients $A_1$, $B_1$, and $C_1$ are found by tensor projection. Writing them as functions of $p$ and the ratio $z=P^0/p$ we find 
\begin{eqnarray}
\label{eq29}
   A_1(p, P^0)\equiv A_1(p,z) &=& m_\infty^2\frac{C_\mathrm{neq}}{\pi^2 p} 
   \left[(5z^2{-}3) Q_0(z) - 5z^2 + \frac{4}{3} \right]  
\notag \\
   B_1(p, P^0) \equiv B_1(p,z) &=& m_\infty^2\frac{C_\mathrm{neq}}{\pi^2 p} 
   \left[ \left(-5z^3 + 6z - \frac{1}{z} \right) Q_0(z) + 5z^3 - \frac{13}{3}z\right]  
\notag \\
   C_1(p, P^0) \equiv C_1(p,z) &=& m_\infty^2\frac{C_\mathrm{neq}}{\pi^2 p} 
   \left[\left(z^2{-}1\right) Q_0(z) - z^2 + \frac{2}{3} \right],
\end{eqnarray}
where $m_\infty^2 = C_F g_s^2 T^2 /4$ is the leading order asymptotic thermal quark mass, and $Q_0(z) = \frac{1}{2} \ln\left(\frac{1{+}z}{1{-}z}\right)$ is the Legendre function of the second kind.

We have now derived all the essential ingredients for the calculation of the photon emission rate in Eq.~(\ref{2.A.1}). In the following two subsections we work out, in turn, the soft and hard contributions to that rate.

%%%%%%%%%%%%%%%%%%%%%%%%%%%%%%%%%%%%%%%%%%%%%
\subsubsection{Soft contribution}
\label{sec2b3}
%%%%%%%%%%%%%%%%%%%%%%%%%%%%%%%%%%%%%%%%%%%%%

Figure~\ref{soft.fig1} shows the Feynman diagrams that need to be evaluated for the soft contribution to the photon emission rate:

%====================== Fig. 2 =================
\begin{figure}[h!]
  \centering
  \includegraphics[width=0.5\linewidth]{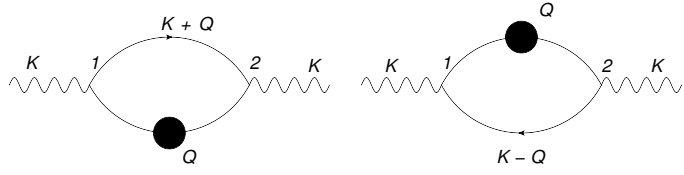} 
  \caption{(12)-component of one-loop photon self-energy with one HTL-resummed quark propagator}
  \label{soft.fig1}
\end{figure}
%=============================================

To leading order in $g_s$, only one of the two quark propagators in the loop requires HTL resummation \cite{Braaten:1989mz,Braaten:1990wp,Kapusta:1991qp,Baier:1991em}, indicated by the blob. The Feynman rules give
\begin{eqnarray}
   i \Pi^\mu_{12\mu} (K) = e^2 \Bigl(\sum_s q_s^2 \Bigr) N_C \!\!\int\!\! \frac{d^4 Q}{(2\pi)^4} 
   \mathrm{Tr}\Bigl[\gamma^\mu i\tilde{S}^\star_{21}(Q) \gamma_\mu iS_{12} (Q{+}K) 
   + \gamma^\mu iS_{21}(Q{-}K) \gamma_\mu i \tilde{S}^\star_{12}(Q)\Bigr],\quad
\label{soft.1}
\end{eqnarray}
where $S_{12}(Q{+}K)$ and $S_{21}(Q{-}K)$ are free quark propagators as in Eq.~(\ref{KMS.2}) and $\tilde{S}^\star_{12}(Q)$ and $\tilde{S}^\star_{21}(Q)$ are hard-loop resummed propagators \cite{Braaten:1990wp,Blaizot:2001nr}:
\begin{equation}
    \tilde{S}^\star_{12(21)}(Q) = \tilde{S}^\star_R(Q) \Sigma_{12(21)}(Q) \tilde{S}^\star_A(Q).
\label{soft.2}
\end{equation}
Using this together with the relations derived in Sec.~\ref{sec2b1} in the hard loop approximation,
\begin{equation}
  \Sigma_{12} (Q) = - \Sigma_{21}(Q) = - i \mathrm{Im}\, \Sigma_R(Q).
\label{soft.3}
\end{equation}
we can rewrite Eq.~(\ref{soft.1}) as
\begin{eqnarray}
   i \Pi^\mu_{12\mu} (K) &=& - e^2 \Bigl(\sum_s q_s^2 \Bigr) N_C \frac{8}{k} f_F(K) 
   \int^{q_\mathrm{cut}} \frac{d^3 q}{(2 \pi)^3} \, \mathrm{Im}\,\bigl(K_\nu \tilde{S}^{\star\nu}_R(Q)\bigr).
\label{soft.7}
\end{eqnarray}
Dynamical quark mass generation through hard loop resummation for the quark propagator is important only in the soft region $Q{\,\ll\,}T$ where the massless bare quark propagator otherwise causes an infrared divergence. On the other hand, only in the soft region $Q{\,\lesssim\,}g_sT{\,\ll\,}T$ is HTL resummation a consistent resummation scheme \cite{Braaten:1989mz}. We therefore introduced here an upper cutoff $q_\mathrm{cut}{\,\sim\,}{\cal O}(g_sT){\,\ll\,}T$ in the $q$ integral and will evaluate the remaining ``hard'' contribution from internal quark momenta $q{\,>\,}q_\mathrm{cut}$ in the following subsection without medium corrections for the internal quark propagator.  

Inserting Eqs.~(\ref{eq1}) and (\ref{soft.8}) into Eq.~(\ref{soft.7}) and linearizing in $\pi_{\alpha \beta}/2(e + \p)$ we obtain
\begin{eqnarray}
    i \Pi^\mu_{12\mu} (K)  &=& - e^2 \Bigl(\sum_s q_s^2 \Bigr) N_C \frac{8}{k} f_{F0}(K) 
    \int^{q_\mathrm{cut}} \frac{d^3 q}{(2 \pi)^3} \biggl[\mathrm{Im} 
    \Bigl\{ \frac{K \cdot Q_0}{Q_0 \cdot Q_0} \Bigr\} 
\notag \\
    && + \frac{\pi_{\alpha \beta}}{2(e + \p)} \biggl(\mathrm{Im}\Bigl\{ -\frac{K_\mu 
       \Sigma_1^{\alpha \beta \mu}}{Q_0 \cdot Q_0} \Bigr\}  
    + \mathrm{Im} \Bigl\{ \frac{K \cdot Q_0}{Q_0 \cdot Q_0} \frac{2 Q_{0\mu} 
       \Sigma_1^{\alpha \beta \mu}}{Q_0 \cdot Q_0} \Bigr\} 
\notag \\
    &&\hspace*{2.5cm}
    + \hat{k}^\alpha \hat{k}^\beta (1{-}f_{F0}(K)) \chi\Bigl(\frac{k}{T}\Bigr) \mathrm{Im} 
    \Bigl\{ \frac{K \cdot Q_0}{Q_0 \cdot Q_0} \Bigr\} \biggr) \biggr],
\label{soft.10}
\end{eqnarray}
where we used the shorthand $Q_0 = Q - \Sigma_0(Q)$. The equilibrium part of the emission rate thus reads
\begin{eqnarray}
   \Gamma_0(K) = - \frac{e^2}{2(2\pi)^3} \Bigl(\sum_s q_s^2 \Bigr) N_C \frac{8 f_{F0}(K)}{k}  
   \int^{q_\mathrm{cut}} \frac{d^3 q}{(2 \pi)^3} \mathrm{Im} \left\{ \frac{K \cdot Q_0}{Q_0 \cdot Q_0} \right\}
\label{soft.11}
\end{eqnarray}
while the viscous correction coefficient is given by
\begin{eqnarray}
   \Gamma^{\alpha\beta}(K) = - \frac{e^2}{2(2\pi)^3} \Bigl(\sum_s q_s^2\Bigr) N_C \frac{8}{k} f_{F0}(K) 
   \!\!\!\!\!&&
   \int^{q_\mathrm{cut}} \!\!\! \frac{d^3 q}{(2 \pi)^3} 
   \bigg[ \hat{k}^\alpha \hat{k}^\beta (1{-}f_{F0}(K)) \chi\Bigl(\frac{k}{T}\Bigr) 
   \mathrm{Im} \Bigl\{ \frac{K \cdot Q_0}{Q_0 \cdot Q_0} \Bigr\}  
\notag \\
   && - \mathrm{Im}\Bigl\{\frac{K_\mu \Sigma_1^{\alpha \beta \mu}}{Q_0 \cdot Q_0}\Bigr\} 
        + \mathrm{Im}\Bigl\{\frac{K \cdot Q_0}{Q_0 \cdot Q_0} \frac{2 Q_{0\mu} \Sigma_1^{\alpha \beta \mu}}
           {Q_0 \cdot Q_0} \Bigr\}\bigg].
\label{soft.12}
\end{eqnarray}

%%%%%%%%%%%%%%%%%%%%%%%%%%%%%%%%%%%%%%%%%%%%%
\subsubsection{Hard contribution}
\label{sec2b4}
%%%%%%%%%%%%%%%%%%%%%%%%%%%%%%%%%%%%%%%%%%%%%

The hard contribution to the photon emission rate can be computed by writing down all two-loop diagrams with bare propagators (\ref{KMS.3}) that contribute to the photon self energy, and computing its imaginary part by applying the finite temperature cutting rules \cite{Majumder:2001iy}. One finds formally the same expression as in thermal equilibrium \cite{Kapusta:1991qp} but with anisotropically modified distribution functions:
\begin{equation}
  k \frac{d R}{d^3 k} = \sum_\mathrm{channels}
  \int_{p, p', k'} \frac{1}{2(2\pi)^3} (2\pi)^4 \delta^{(4)}(P{+}P'{-}K{-}K') 
  \vert \mathcal{M} \vert^2 f(P) f(P') (1{\pm}f(K')),
\label{hard.1}
\end{equation}
where $\int_p$ is a shorthand notation for $\frac{1}{(2\pi)^3} \int \frac{d^3p}{2 P^0}$ (all incoming and outgoing particles are on-shell and massless). Note that the same expression, with modified matrix elements, is used in the kinetic approach~\cite{Weldon:1983jn} discussed in the following subsection.

There are two contributing processes, (anti-)quark-gluon Compton scattering $q{+}g\to q{+}\gamma$, $\bar{q}{+}g\to \bar{q}{+}\gamma$, and quark-antiquark annihilation $q{+}\bar{q}\to\gamma{+}g$ (see Fig.~\ref{fig1}). For Compton scattering $\vert \mathcal{M} \vert^2 \propto -\frac{s}{t} - \frac{t}{s}$ while for pair annihilation $\vert \mathcal{M} \vert^2 \propto \frac{u}{t}$. We treat the phase space integrals as done in \cite{Arnold:2001ms}, handling the three infrared divergent $t$-channel diagrams in Fig.~\ref{fig1} together and the finite $s$-channel diagram separately. In the $t$-channel part the change of variables $Q=P{-}K$ facilitates implementation of the phase space cut $q{\,>\,}q_\mathrm{cut}$ to excise the infrared divergence in a manner that perfectly complements the calculation of the soft contribution in the preceding subsection \cite{Kapusta:1991qp}. No such cut is needed for the $s$-channel diagram.

%===================  Fig. 3 ====================
\begin{figure}[h!]
  \centering
  \includegraphics[width=0.4\linewidth]{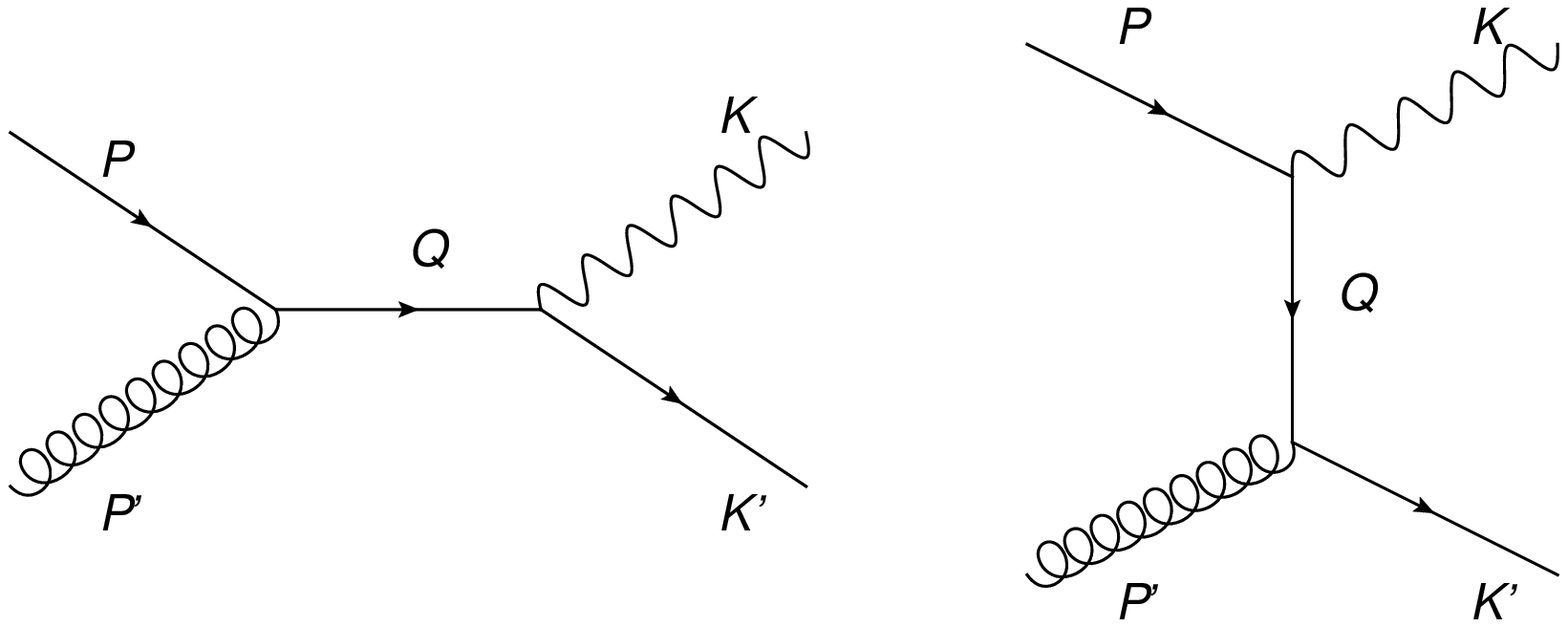} \\
  \includegraphics[width=0.4\linewidth]{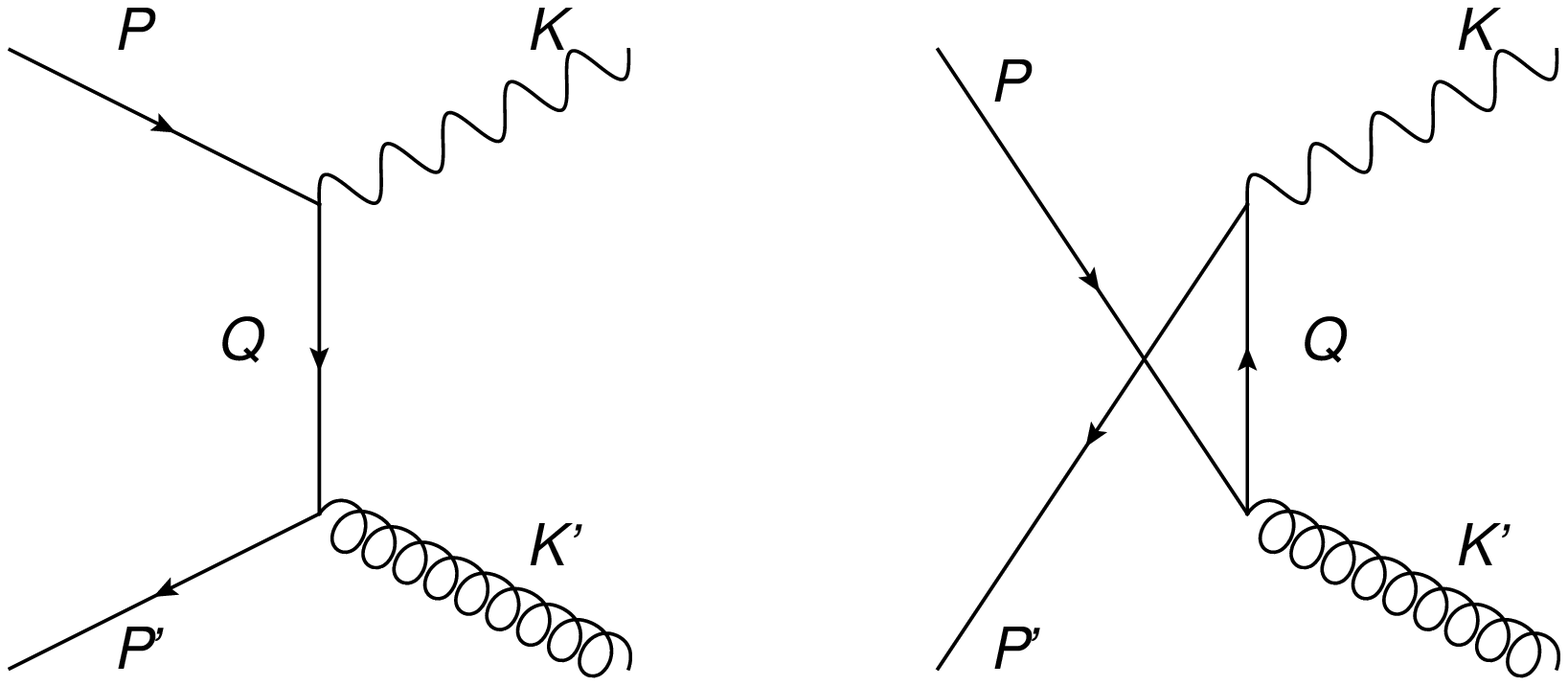}
  \caption{Compton scattering and pair annihilation. Compton scattering can involve gluons scattering 
  off quarks (shown) or antiquarks (not shown).}
  \label{fig1}
\end{figure}
%=======================================

The details of the calculation are presented in Appendix~\ref{appHard}. The final expression for the viscous correction to the photon production rate is very similar to the corresponding ideal rate in \cite{Arnold:2001ms}, in the sense that it is a multidimensional integral over the same variables and with the same kinematic limits, but with a modified integrand.

One should note that strictly speaking this calculation is only valid for internal quark momenta $q{\,\sim\,}{\cal O}(T){\,\gg\,}g_sT$ \cite{Braaten:1990wp,Kapusta:1991qp,Baier:1991em} whereas the soft part, Eqs. (\ref{soft.11}) and (\ref{soft.12}), is valid only for $q{\,\sim\,}{\cal O}(g_sT){\,\ll\,}T$. In Sec.~\ref{sec3a} we will explore to what extent there exists a ``window of insensitivity'' $g_sT{\,\ll\,}q_\mathrm{cut}{\,\ll\,}T$ where both approximations are simultaneously valid and can be matched to each other without strong dependence on the cutoff $q_\mathrm{cut}$. 

%%%%%%%%%%%%%%%%%%%%%%%%%%%%%%%%%%%%%%%%%%%%%%%%%%%%
\subsection{Kinetic Approach}
\label{sec2c}
%%%%%%%%%%%%%%%%%%%%%%%%%%%%%%%%%%%%%%%%%%%%%%%%%%%%

Photon emission rates for $2\to 2$ scattering processes can also be calculated in the kinetic approach sketched in Eq.~(\ref{2.A.3}), involving a sum of terms corresponding to the Compton scattering and pair annihilation channels shown in Fig.~\ref{fig1}. In the equilibrated case this was shown to be equivalent to the diagrammatic approach up to subleading corrections in $g_s$ \cite{Kapusta:1991qp,Baier:1991em,Majumder:2001iy}.
(In fact, the equivalence can be extended to the full leading-order rate by suitably modifying the structure of the collision term in the kinetic description \cite{Arnold:2002zm}.) In Compton scattering and pair annihilation, logarithmic infrared divergences will be generated in the $t$ and $u$ channels if one uses scattering matrix elements computed with free fermion propagators for the internal exchanged quark. This infrared sensitivity is cut off by using the retarded hard loop resummed self-energy $\Sigma(Q)$ for the internal quark propagator in these matrix elements.

The $s$-channel processes are free from infrared singularities and do not require HTL resummation. In fact, using HTL resummed internal quark propagators in the $s$-channel process would cause problems because the collision integral integrates over a kinematic domain where the time-like virtual quark goes on-shell and becomes a long-lived quasi-particle excitation in the medium \cite{Arnold:2002zm}. This is kinematically allowed even with massless external particles. In \cite{Arnold:2002zm} such processes are denoted as ``2 $\rightarrow$ 1 joining'', and the authors of \cite{Arnold:2002zm} point out that they are automatically included in an improved treatment that extends the validity of the calculation from leading logarithmic to full leading order in $g_s$, by properly including LPM effects. Including a fraction of these effects  separately in the $2\to2$ $s$-channel collisions by using HTL resummed internal propagators is not a consistent procedure and, when combined with a consistent LPM treatment \cite{Arnold:2001ms,Arnold:2002ja}, would amount to double counting. For these reasons we use here in the kinetic approach matrix elements that include HTL resummed internal quark propagators in the $u$ and $t$ channels, but not in the $s$-channel. 

We note that in the $u$ and $t$ channels HTL resummation is required for consistency {\em at leading order} in the soft exchange region but not for hard scatterings where it contributes only at {\em next-to-leading order} in $g_s$. As mentioned earlier, using the HTL resummed propagators everywhere is not a consistent approximation scheme, but the inconsistencies are restricted to subleading order in $g_s$. In the diagrammatic approach  described in the preceding subsection, we use free internal quark propagators for hard collisions, consistently matched to matrix elements using resummed internal propagators in the soft region. In the kinetic approach described in the present section, we use HTL resummed matrix elements for the entire kinematic range. The difference amounts to different prescriptions for a partial resummation of higher order terms that are subleading in $g_s$. For sufficiently small $g_s$, both approaches are expected to yield identical results; for moderate values of $g_s$, the differences between the approaches can be taken as a (rough) indicator for the theoretical uncertainties associated with the higher order corrections to our calculation. 

The matrix element for Compton scattering in QGP can be written as
\begin{eqnarray}
   \sum_\mathrm{spin} \sum_\mathrm{color} \vert M_\mathrm{Comp} \vert^2_\mathrm{eq} 
   &=& e^2 g^2 (t^a t^a) 
\notag \\
   &\times& \bigg\{ \frac{16}{\vert Q \cdot Q \vert^2} \Bigl(2 \mathrm{Re}\bigl[(K' \cdot Q)(P \cdot Q^*)\bigr] 
                  - (K' \cdot P) (Q \cdot Q^*)\Bigr) 
\notag \\
   &-& 64 (K' \cdot P) \mathrm{Re} \Bigl[ \frac{Q \cdot Q^{'*}}{(Q \cdot Q)(Q^{'*} \cdot Q^{'*})} \Bigr] 
\notag \\
   &+&\frac{16}{\vert Q' \cdot Q' \vert^2} 
   \Bigl(2 \mathrm{Re}\bigl[(K' \cdot Q')(P \cdot Q^{'*})\bigr] - (K' \cdot P) (Q' \cdot Q^{'*})\Bigr)\bigg\},
\label{kinetic.1}
\end{eqnarray}
where $Q^\mu = P^\mu + P'^\mu $ and $Q'^\mu = P^\mu - K^\mu - \Sigma_R^\mu(P{-}K)$. For pair annihilation we have similarly
\begin{eqnarray}
  \sum_\mathrm{spin} \sum_\mathrm{color} \vert M_\mathrm{pair} \vert^2_\mathrm{eq} 
  &=& e^2 g^2 (t^a t^a) 
\notag \\
   &\times& \bigg\{ \frac{16}{\vert Q' \cdot Q' \vert^2} \Bigl(2 \mathrm{Re}
   \bigl[(P' \cdot Q')(P \cdot Q^{'*})\bigr] - (P' \cdot P) (Q' \cdot Q^{'*})\Bigr) 
\notag \\
   &-& 64 (P' \cdot P) \mathrm{Re} \Bigl[ \frac{\tilde{Q} \cdot Q^{'*}}{(\tilde{Q} \cdot \tilde{Q})
               (Q^{'*} \cdot Q^{'*})} \Bigr] 
\notag \\
   &+&\frac{16}{\vert \tilde{Q} \cdot \tilde{Q} \vert^2} \Bigl(2 \mathrm{Re} 
   \bigl[(P' \cdot \tilde{Q})(P \cdot \tilde{Q}^{*})\bigr] - (P' \cdot P) (\tilde{Q} \cdot \tilde{Q}^{*})\Bigr)\bigg\},
\label{kinetic.2}
\end{eqnarray}
where $Q'^\mu = P^\mu - K^\mu - \Sigma_R^\mu(P{-}K)$ and $\tilde{Q}^\mu = P^\mu - K'^\mu - \Sigma_R^\mu(P{-}K')$. The matrix elements for both channels involve the retarded quark self-energy $\Sigma_R^\mu$ calculated in Eqs.~(\ref{selfenergy.4})-(\ref{eq29}). 

%%%%%%%%%%%%%%%%%%%%%%%%%%%%%%%%%%%%%%%%%%%%%
\section{Results and Discussions}
\label{sec3}
%%%%%%%%%%%%%%%%%%%%%%%%%%%%%%%%%%%%%%%%%%%%%

In this section we compute and graph the photon emission rates calculated with the diagrammatic and the kinetic approaches and compare the two approaches. By default, we employ $\chi\left(\frac{p}{T}\right)=(p/T)^2$ for the momentum dependence of $\delta f$ in Eq. (\ref{eq1}), i.e. we set $\lambda{\,=\,}2$. The $\lambda$ dependence of the photon emission rates is studied at the end of this section.

For completeness we also compare our rates with two other approaches currently on the market: the $2\to 2$ part of the ideal rate from AMY \cite{Arnold:2001ms}, and the viscous calculation using the forward-scattering dominance approximation (FSDA) presented in Ref.~\cite{Dusling:2009bc}. 

The calculation from AMY is formally equivalent to the diagrammatic approach described in this paper. Our treatment differs from theirs, however, in the way we splice together the soft and hard contributions. As we explain in the following section, this leads to differences in the total rate when $g_s$ is not small.

The use of the forward-scattering dominance approximation in Ref.~\cite{Dusling:2009bc} strongly simplifies the photon rate calculation compared with the full approach used in the present paper.\footnote{
The ansatz used in Ref.~\cite{Dusling:2009bc} for the momentum anisotropy differs slightly from ours, but the formula (Eq.~(5) in~\cite{Dusling:2009bc})
\begin{equation}
k \frac{d R}{d^3 k}=\frac{e^2 g_s^2 \left(\sum_s q_s^2 \right)}{\pi(2\pi)^3} f(K)\, T^2 \ln \left[ \frac{3.7388\, k}{g_s^2 T} \right]
\end{equation}
can be straightforwardly adapted to our case by replacing the ansatz in \cite{Dusling:2009bc} for $f(K)$ by our Eq.~(\ref{eq1}). The results of doing so are labelled as ``Ref.~\cite{Dusling:2009bc}''
in the figures below.
}
Comparing the results from Ref.~\cite{Dusling:2009bc} with our full calculation allows to better understand the region of validity and accuracy of that approach.

To compare the different approaches we plot the dimensionless equilibrium rates and viscous correction coefficients $\tilde\Gamma_0$ and $\tilde\Gamma_1$ in Eq.~(\ref{eq2}), as well as their ratio $\tilde \Gamma_1/\tilde\Gamma_0{\,=\,}\Gamma_1/\Gamma_0$, as functions of $k/T$ for selected values of the parameters $g_s$ (resp. $\alpha_s{\,=\,}g_s^2/(4\pi)$) and $\lambda$, and of the cutoff momentum $q_\mathrm{cut}/T$ that separates the hard and soft scattering domains in the diagrammatic approach.

%%%%%%%%%%%%%%%%%%%%%%%%%%%%%%%%%%%%%%%%%%%%%
\subsection{Cut-off dependence in diagrammatic approach}
\label{sec3a}
%%%%%%%%%%%%%%%%%%%%%%%%%%%%%%%%%%%%%%%%%%%%%

Recall that the cutoff introduced in the diagrammatic approach is artificial: the hard scattering sector where medium corrections to the matrix elements are negligible should match smoothly to the soft scattering region where HTL resummation of the self-energy is essential to regulate the infrared logarithmic divergence. Formally the value of the cutoff $q_\mathrm{cut}$ should satisfy $g_s T{\,\ll\,}q_\mathrm{cut}{\,\ll\,}T$. Physically, the final photon emission rates should be completely insensitive to this artificial cut-off, while in practice this means that there should be a range of values for $q_\mathrm{cut}$ between $g_s T$ and $T$ for which the rate is largely insensitive to $q_\mathrm{cut}$. On the other hand it also means that this cutoff independence should quickly evaporate when $g_s T{\,\gtrsim\,}T$, i.e. when the soft and hard scales overlap. Both of these issues are investigated in this subsection. To this end, we explore the behavior of the thermal photon rates for two values of $g_s$: $g_s{\,=\,}0.01{\,\ll\,}1$ (corresponding to weak coupling $\alpha_s{\,=\,}g_s^2/(4\pi){\,\approx\,}8\times10^{-6}$ and well-separated soft and hard scales) and the more realistic value (for RHIC and LHC applications) $g{\,=\,}2$ (corresponding to moderately strong coupling $\alpha_s{\,\simeq\,}0.3$ and overlapping soft and hard scales).

%======================= Fig. 4 =========================================
\begin{figure}[h]
\centering
\begin{tabular}{cc}
\includegraphics[width=0.48\linewidth]{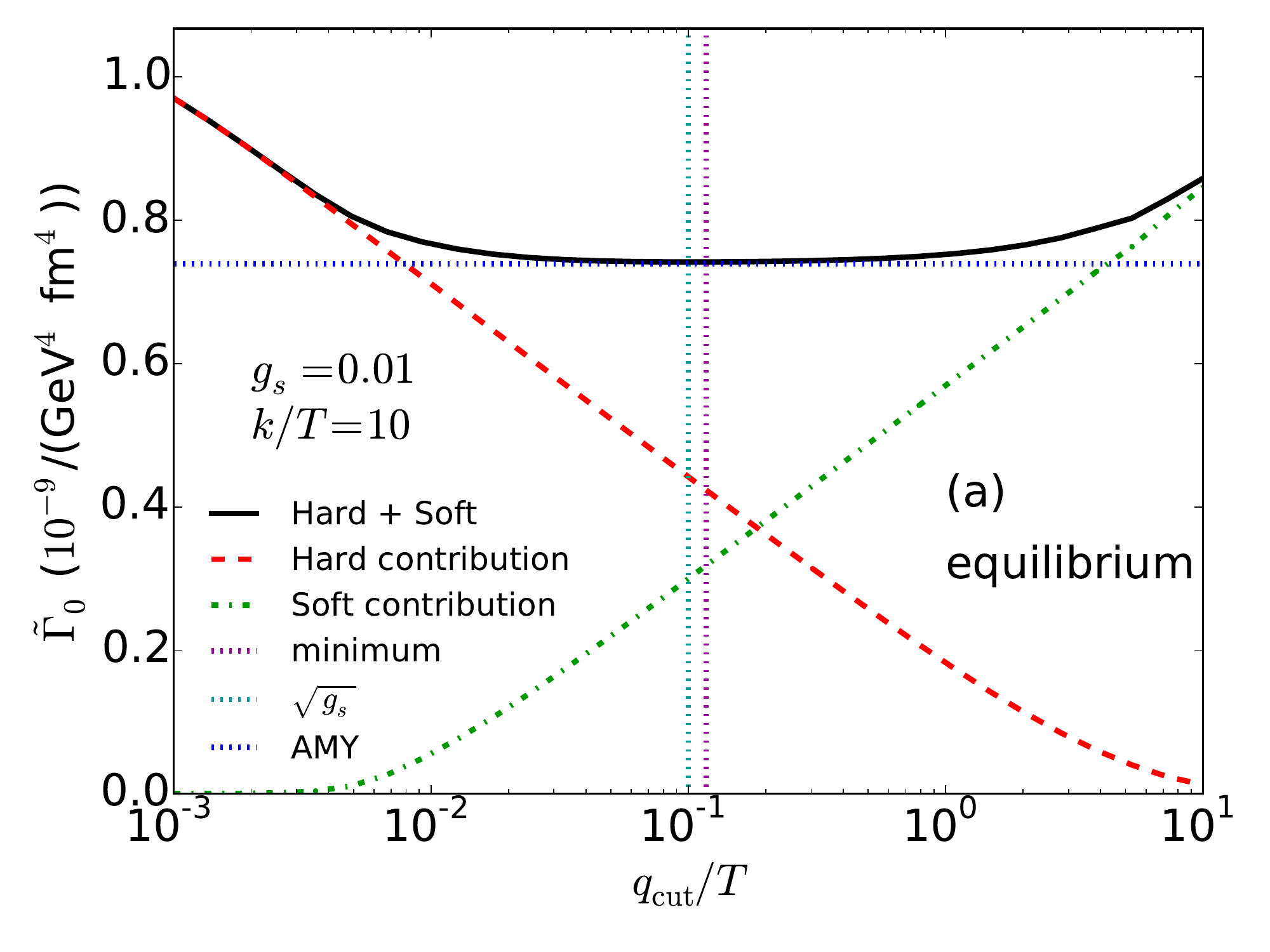} &
\includegraphics[width=0.48\linewidth]{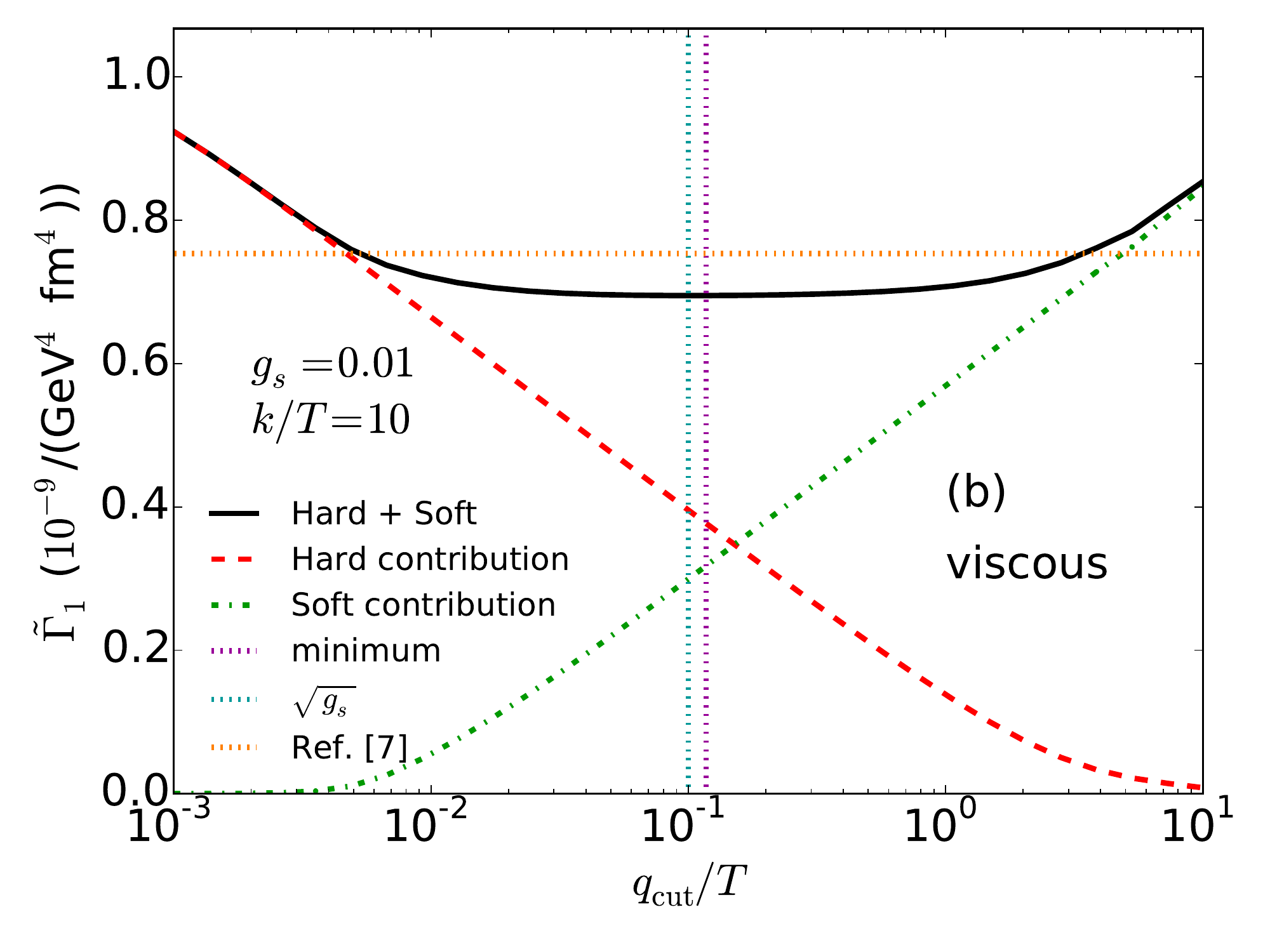} \\
\includegraphics[width=0.48\linewidth]{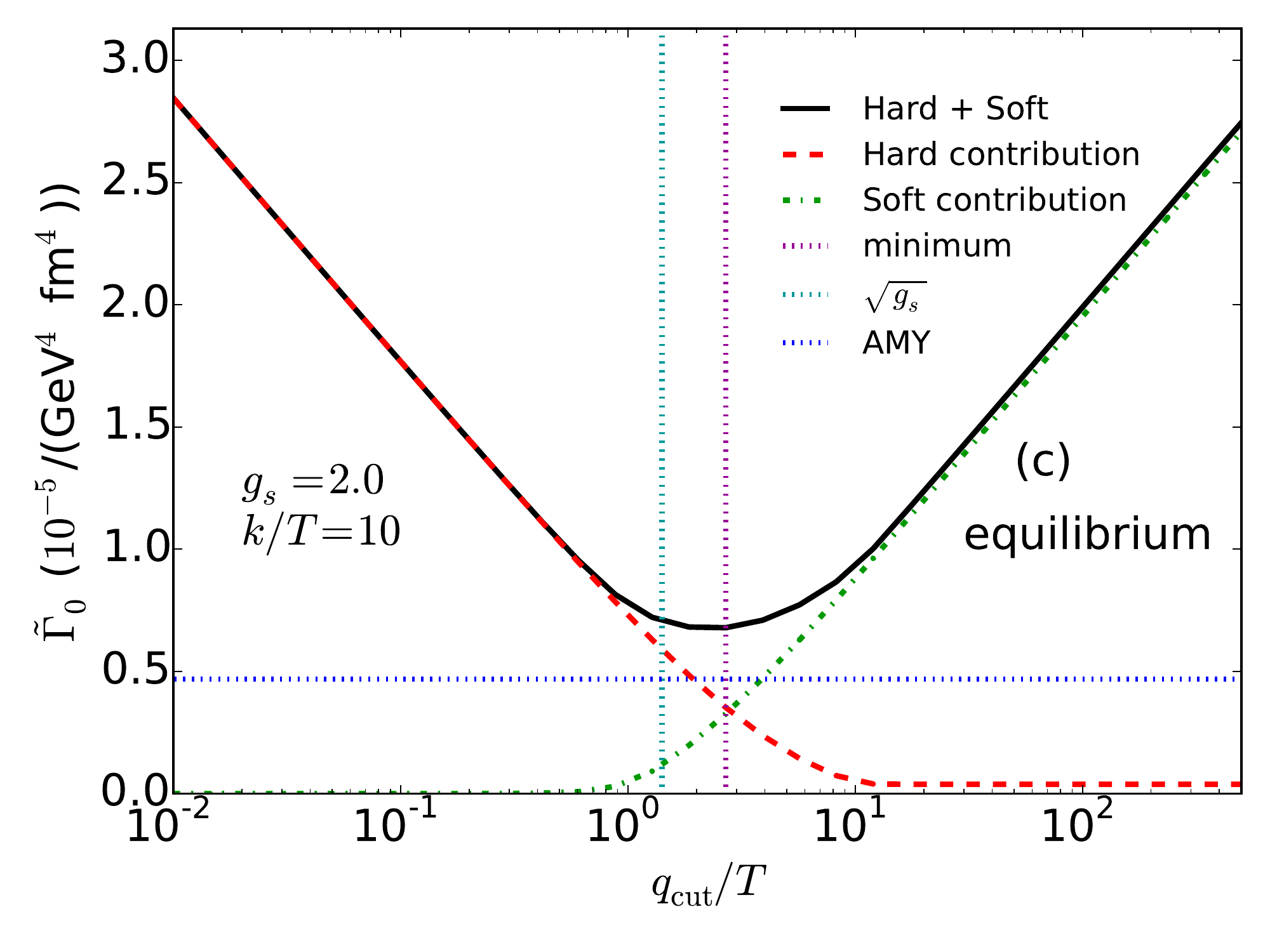} &
\includegraphics[width=0.48\linewidth]{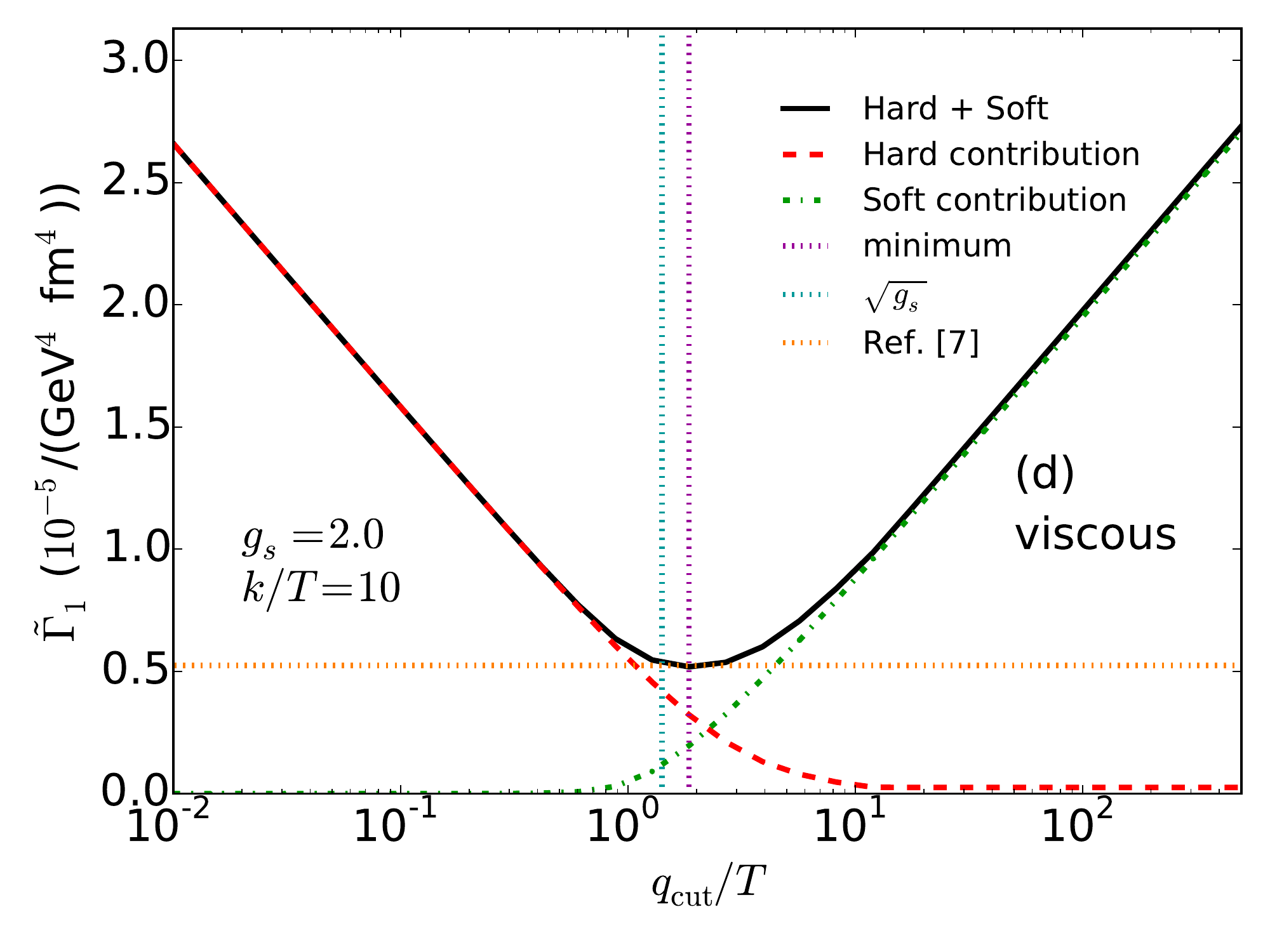}
\end{tabular}
\caption{(Color online) Cutoff dependence of the normalized equilibrium rate $\tilde\Gamma_0$ (a,c) and viscous correction coefficient $\tilde\Gamma_1$ (b,d) from the diagrammatic approach at $k/T = 10$, for two values of the strong coupling constant, $g_s = 0.01$ (a,b) and $g_s = 2.0$ (c,d). Horizontal dotted lines indicate the value from the AMY parametrization \cite{Arnold:2001ms} in (a,c) and from Ref.~\cite{Dusling:2009bc} in (b,d). Vertical dotted lines indicate the positions of the minima of the numerical curves and of $q_\mathrm{cut}/T = \sqrt{g_s}$, respectively. See text for discussion. 
}
\label{fig6}
\end{figure}
%======================================================================

%======================= Fig. 5 =========================================
\begin{figure}[h]
\centering
\begin{tabular}{cc}
\includegraphics[width=0.48\linewidth]{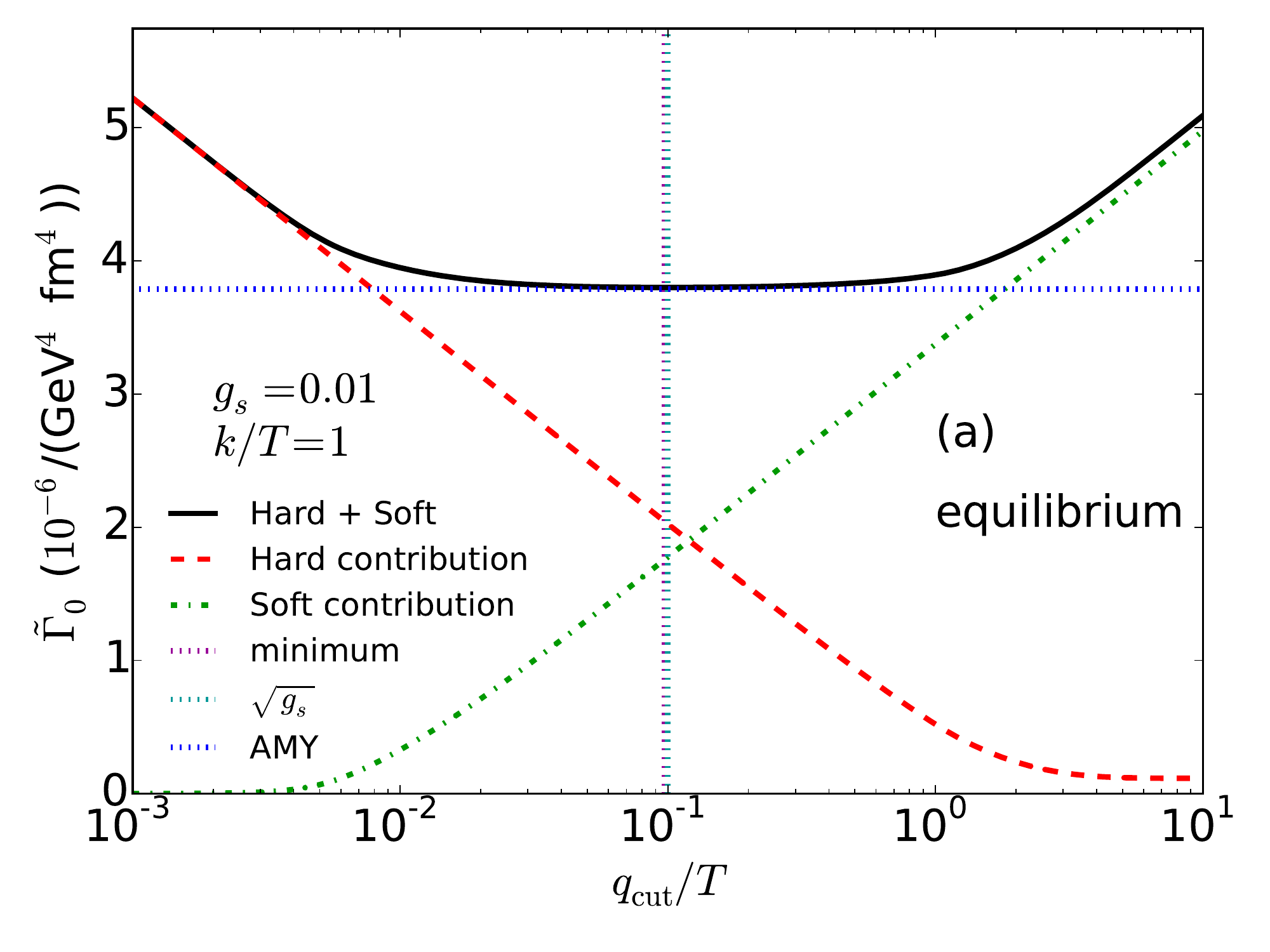} &
\includegraphics[width=0.48\linewidth]{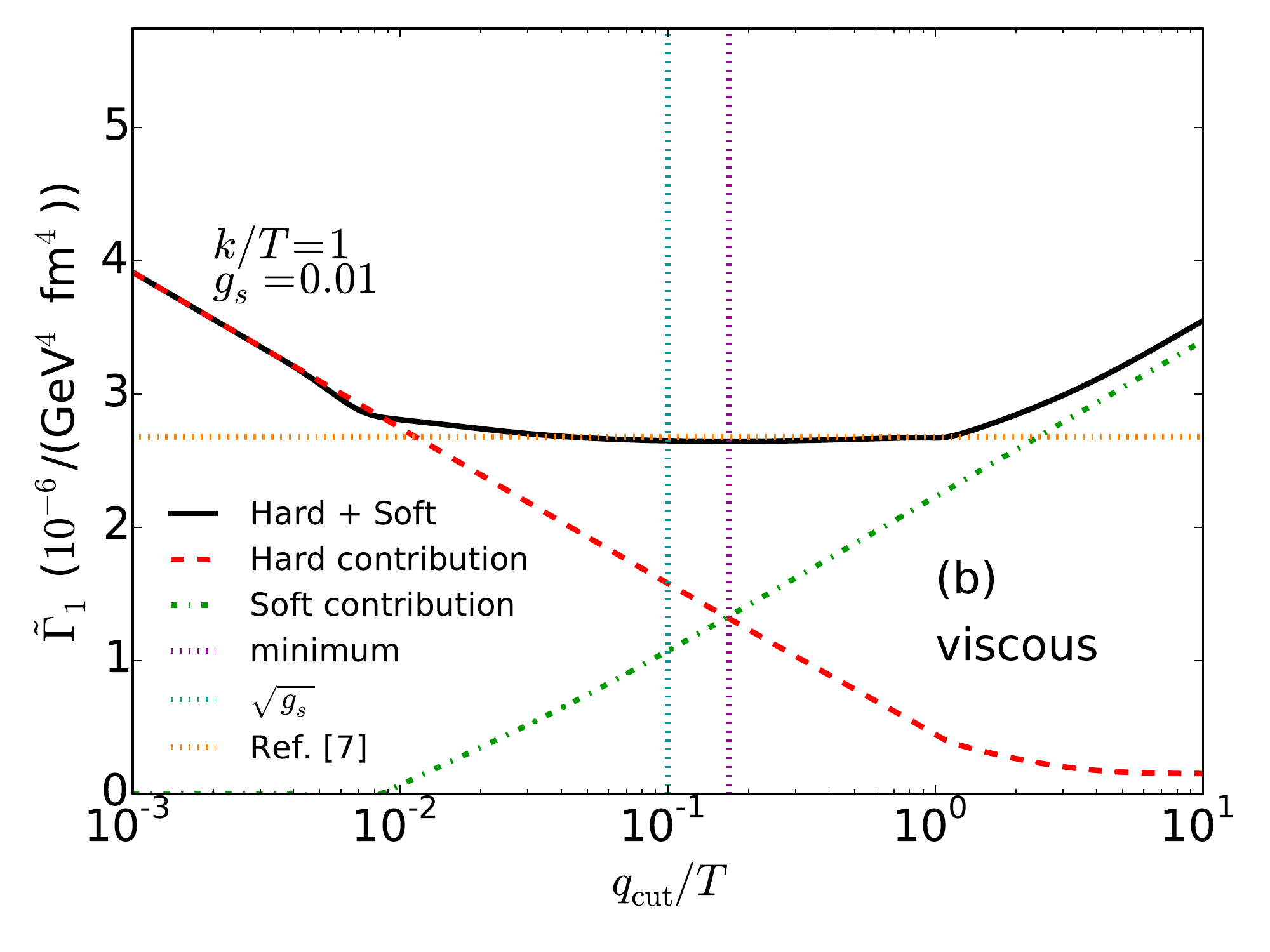} \\
\includegraphics[width=0.48\linewidth]{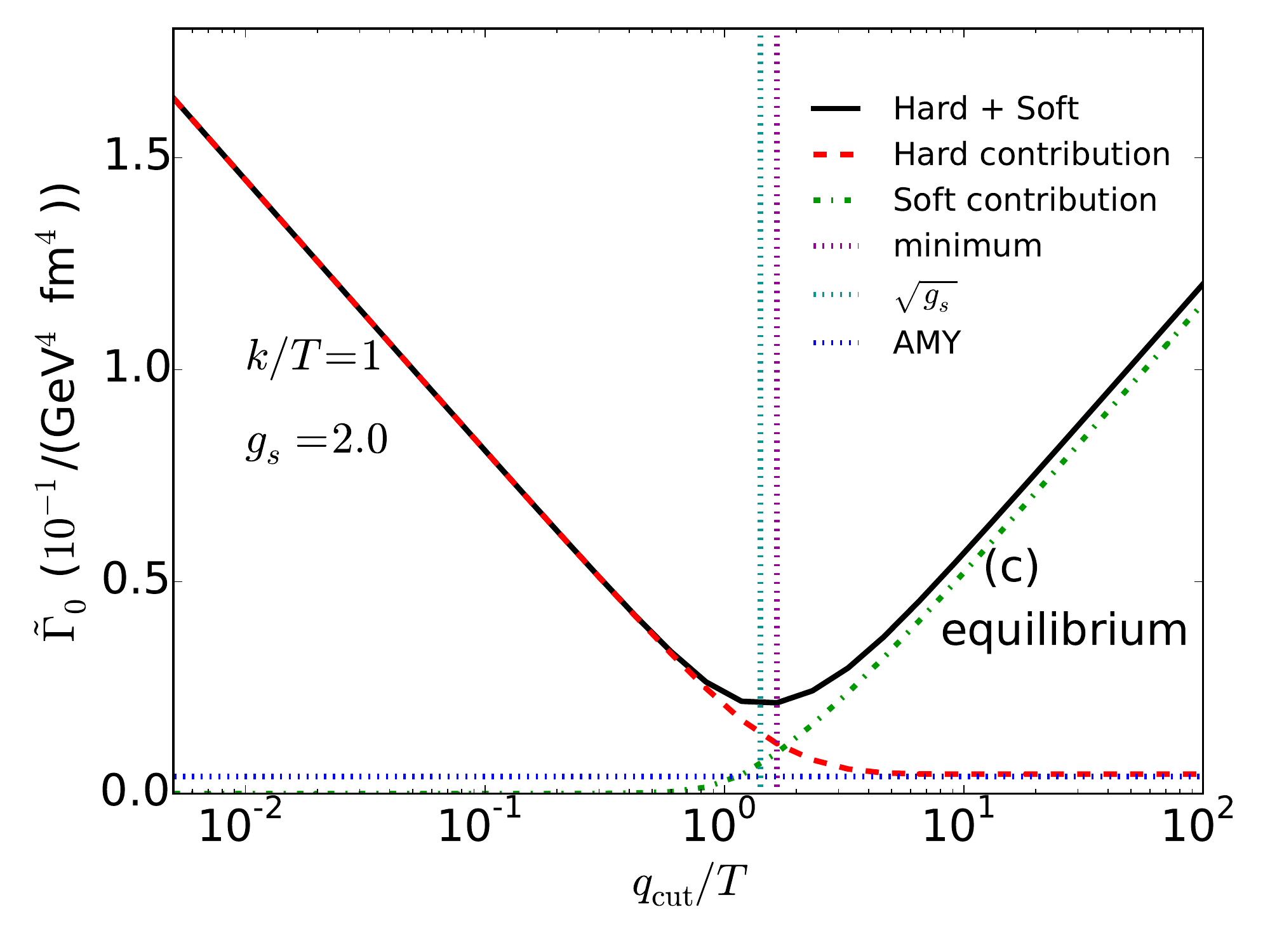} &
\includegraphics[width=0.48\linewidth]{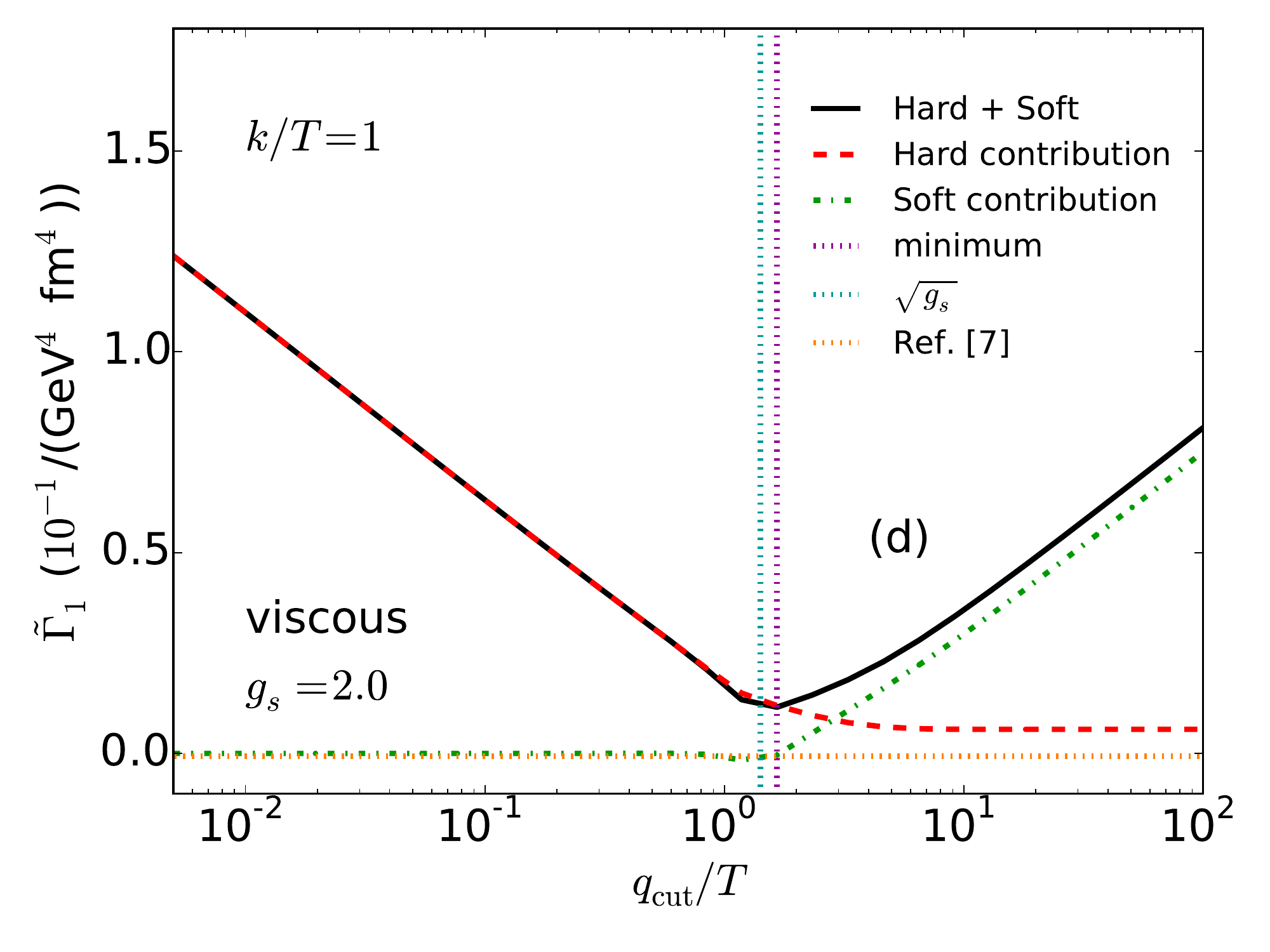}
\end{tabular}
\caption{(Color online) Same as Fig.~\ref{fig6}, but for softer photons at $k/T{\,=\,}1$.}
\label{fig6prime}
\end{figure}
%======================================================================

In Fig.\,\ref{fig6} we plot the scaled (dimensionless) photon emission rate $\tilde\Gamma_0$ and the viscous correction coefficient $\tilde\Gamma_1$ from the diagrammatic approach at $k/T = 10$, as a function of the scaled cutoff momentum $q_\mathrm{cut}/T$.  Figs.~\ref{fig6}a,b show that for weak coupling, $g_s = 0.01$, both $\tilde\Gamma_0$ and $\tilde\Gamma_1$ exhibit a wide plateau that extends roughly from $q_\mathrm{cut}/T=g_s$ to $q_\mathrm{cut}/T=1$, with a shallow minimum near $q_\mathrm{cut}/T=\sqrt{g_s}$. In the plateau region, the total rates (soft+hard) are practically cutoff-independent.

For larger coupling this window of insensitivity shrinks, and for $g_s=2$ (panels (c) and (d)) it has disappeared. Still, for both $\tilde\Gamma_0$ and $\tilde\Gamma_1$, the sum of soft and hard scattering contributions is still minimal near $q_\mathrm{cut}/T=\sqrt{g_s}$. In the following sections, we adopt the sum of the hard and soft contributions at $q_\mathrm{cut}/T=\sqrt{g_s}$ as our estimate for $\Gamma_{0,1}$ from the diagrammatic approach.

This prescription agrees with the one adopted in \cite{Schenke:2006yp} but not with the approach taken by AMY in \cite{Arnold:2001ms}. AMY start from the observation that for sufficiently small coupling the $q_\mathrm{cut}/T$ dependences of the soft and hard contributions to the thermal photon rate must cancel exactly, and that in the asymptotic regions ($q_\mathrm{cut}/T{\,\gg\,}1$ for the soft contribution, $q_\mathrm{cut}/T{\,\ll\,}1$ for the hard one) the cutoff dependences of both contributions are linear in $\ln(q_\mathrm{cut}/T)$ (with opposite slopes). They then eliminate the $\ln(q_\mathrm{cut}/T)$ dependence of the total rate by adding these two asymptotic logarithmic terms; this leads to the horizontal dotted lines in Figs.~\ref{fig6}a,c and \ref{fig6prime}a,c. We see on Figs.~\ref{fig6}c and \ref{fig6prime}c, however, that for $g_s{\,=\,}2$ the cutoff dependences of the hard and soft contributions to the rate are no longer linear in $\ln(q_\mathrm{cut}/T)$ in the region $q_\mathrm{cut}/T{\,\sim\,}1$ where the soft and hard contributions should be matched. For moderately strong coupling, evaluating both contributions numerically and adding them as we do here therefore gives a larger result than the one obtained by AMY. These observations hold for both low ($k/T{\,=\,}1$, Fig.~\ref{fig6prime}) and high ($k/T{\,=\,}10$, Fig.~\ref{fig6}) photon energies.

Ref.~\cite{Dusling:2009bc} does not have a cut-off dependence either: the $q_\mathrm{cut}/T$ dependence of the hard contribution, as evaluated with the forward dominance scattering approximation, is canceled against the asymptotic cut-off dependence of the soft part. 
We note that for the viscous correction, Ref.~\cite{Dusling:2009bc} makes another approximation that affects the cut-off dependence: terms that were found~\cite{Dusling:2009bc} to be subleading in $\log(g_s)$ are neglected in both the soft and hard part of the viscous correction.
Neglecting these terms simplifies the evaluation of the viscous correction, but also have the side effect of removing terms that would have had a cut-off dependence.

Fig.~\ref{fig6}b shows that, even for weak coupling $g_s{\,\ll\,}1$ where our calculations show a wide window of insensitivity of $\tilde\Gamma_1$ to $q_\mathrm{cut}/T$, this approximation leads to somewhat larger $\tilde\Gamma_1$ values than our estimate. At stronger coupling ($g_s{\,=\,}2$, Fig.~\ref{fig6}d), the various approximations of Ref.~\cite{Dusling:2009bc} accidentally cancel each other, yielding a result very close to our diagrammatic calculation. At smaller photon energy $k/T{\,=\,}1$ this cancellation no longer happens at $g_s{\,=\,}2$ (Fig.~\ref{fig6prime}d), instead it has moved to $g_s{\,\simeq\,}0.01$ (Fig.~\ref{fig6prime}b). A more detailed comparison of the rates is made in the next section.

%%%%%%%%%%%%%%%%%%%%%%%%%%%%%%%%%%%%%%%%%%%%%
\subsection{Rate comparison}
\label{sec3b}
%%%%%%%%%%%%%%%%%%%%%%%%%%%%%%%%%%%%%%%%%%%%%

We now compare our results from the diagrammatic approach with the kinetic approach, along with the results from Ref.~\cite{Dusling:2009bc} and AMY when relevant. As stated above, we use $q_\mathrm{cut} = \sqrt{g_s}T$ for the momentum cutoff in the diagrammatic approach since this value is generally close to the region of minimal cutoff dependence. 
%
%================================= Fig. 6 ==========================================
\begin{figure}[h]
\centering
\begin{tabular}{cc}
\includegraphics[width=0.48\linewidth]{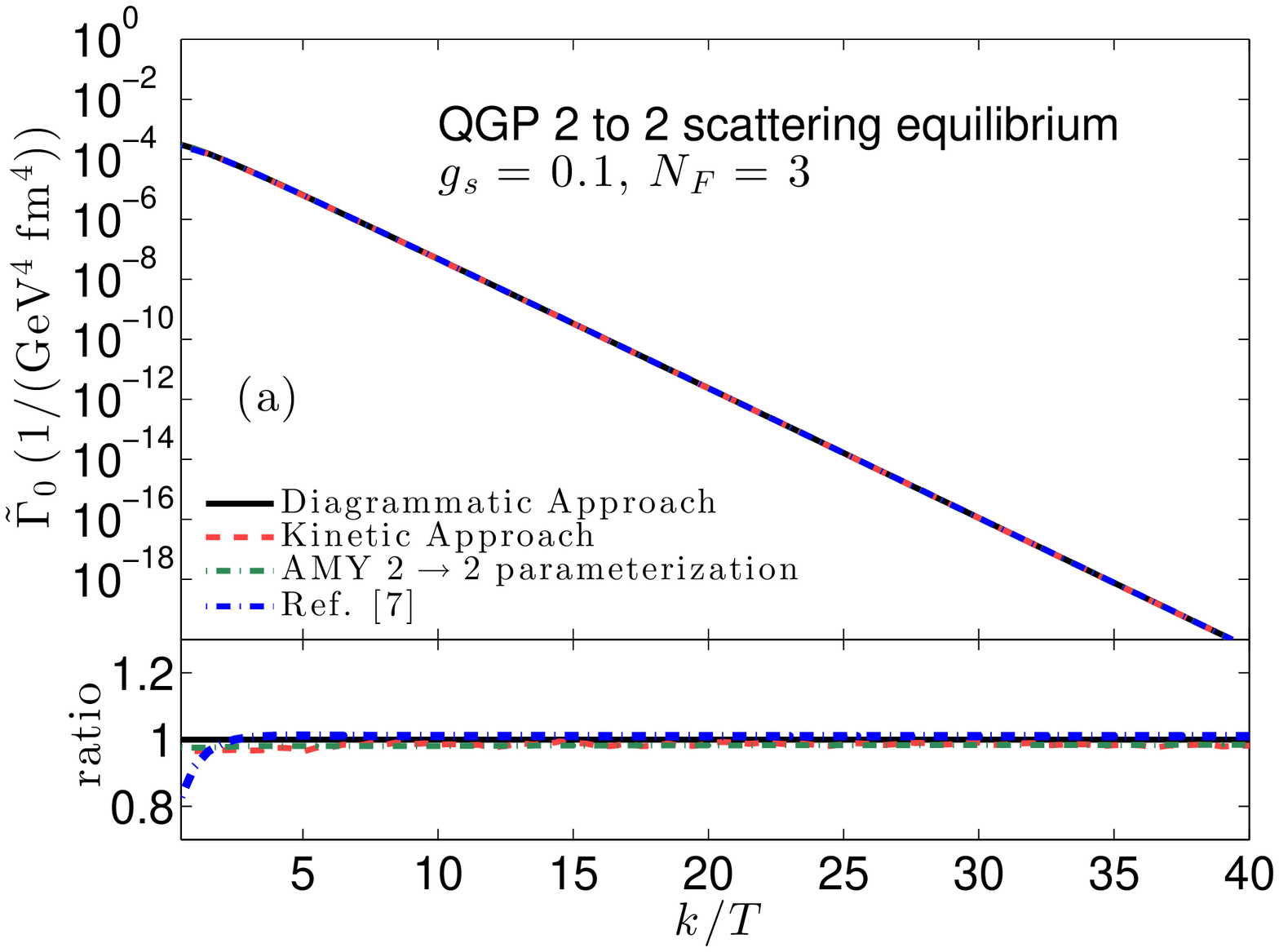} &
\includegraphics[width=0.48\linewidth]{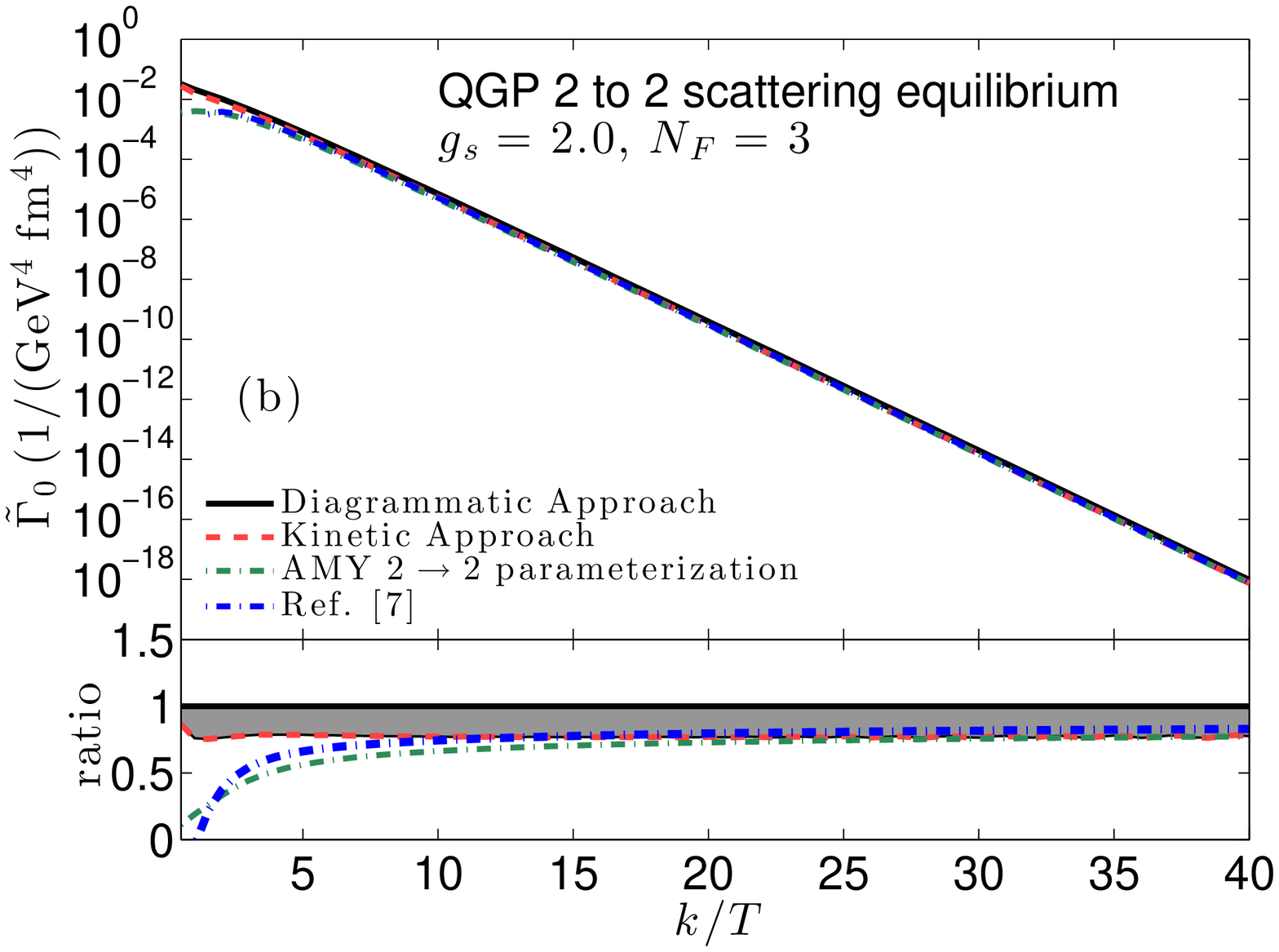}
\end{tabular}
\caption{(Color online) The temperature-scaled equilibrium photon emission rate $\tilde\Gamma_0$ as a function of $k/T$ for relatively weak  ($g_s{\,=\,}0.1$, (a)) and moderately strong coupling ($g_s{\,=\,}2$, (b)). Results are shown for the diagrammatic approach, the kinetic approach, AMY's parametrization \cite{Arnold:2001ms}, and for Ref.~\cite{Dusling:2009bc} as labelled. In the lower panels we show the ratio between these rates and the one from the diagrammatic approach on a linear scale. The gray band is a lower bound on the uncertainty of higher order corrections in $g_s$ to the diagrammatic and kinetic approaches.}
\label{fig4}
\end{figure}
%=================================================================================
%
Figures~\ref{fig4} and \ref{fig5} show the scaled equilibrium rate $\tilde\Gamma_0$ and the viscous correction coefficient $\tilde\Gamma_1$ as functions of $k/T$, for two values of the coupling constant, $g_s{\,=\,}0.1$ (a) and $g_s{\,=\,}2$ (b). For $g_s{\,=\,}0.1$ in Fig.\,\ref{fig4}a, the equilibrium photon emission rates from all four approaches are found to agree with each other very well. The difference between our numerical results and AMY's parametrization is within 2\%. The result from Ref.~\cite{Dusling:2009bc} deviates from the others only for $k/T{\,<\,}1$.

For $g_s{\,=\,}2.0$ (Fig.~\ref{fig4}b), the thermal equilibrium rates from the four approaches show similar $k/T$ dependences for $k/T > 5$. However the diagrammatic approach shows a systematically higher normalization than the three other calculations. In particular, the difference between the kinetic and diagrammatic approaches, which amounts to about 25\% independent of $k/T$ in the range $1{\,<\,}k/T{\,<40\,}$, is a manifestation of a different partial resummation of higher order corrections in the two methods, as discussed at the beginning of Sec.~\ref{sec2c}. It can be taken as an indicator (or more precisely, a lower limit) of the systematic uncertainty of our calculation when extrapolating the result to moderately large coupling $\alpha_s{\,=\,}0.3$. The difference between the AMY parametrization and our diagrammatic approach is due to the different treatment to the cutoff dependence. 

For $k/T{\,>\,}20$, the thermal equilibrium rates from both AMY and Ref.~\cite{Dusling:2009bc} agree within a few percent with our kinetic theory results. However, both calculations start to deviate from the kinetic approach with full matrix elements for $k/T{\,<\,}10$ where higher order corrections presumably become increasingly important. The result from Ref.~\cite{Dusling:2009bc} actually goes negative for $k/T{\,<\,}1$.

%============================== Fig. 7 ============================================
\begin{figure}[h]
\centering
\begin{tabular}{cc}
\includegraphics[width=0.48\linewidth]{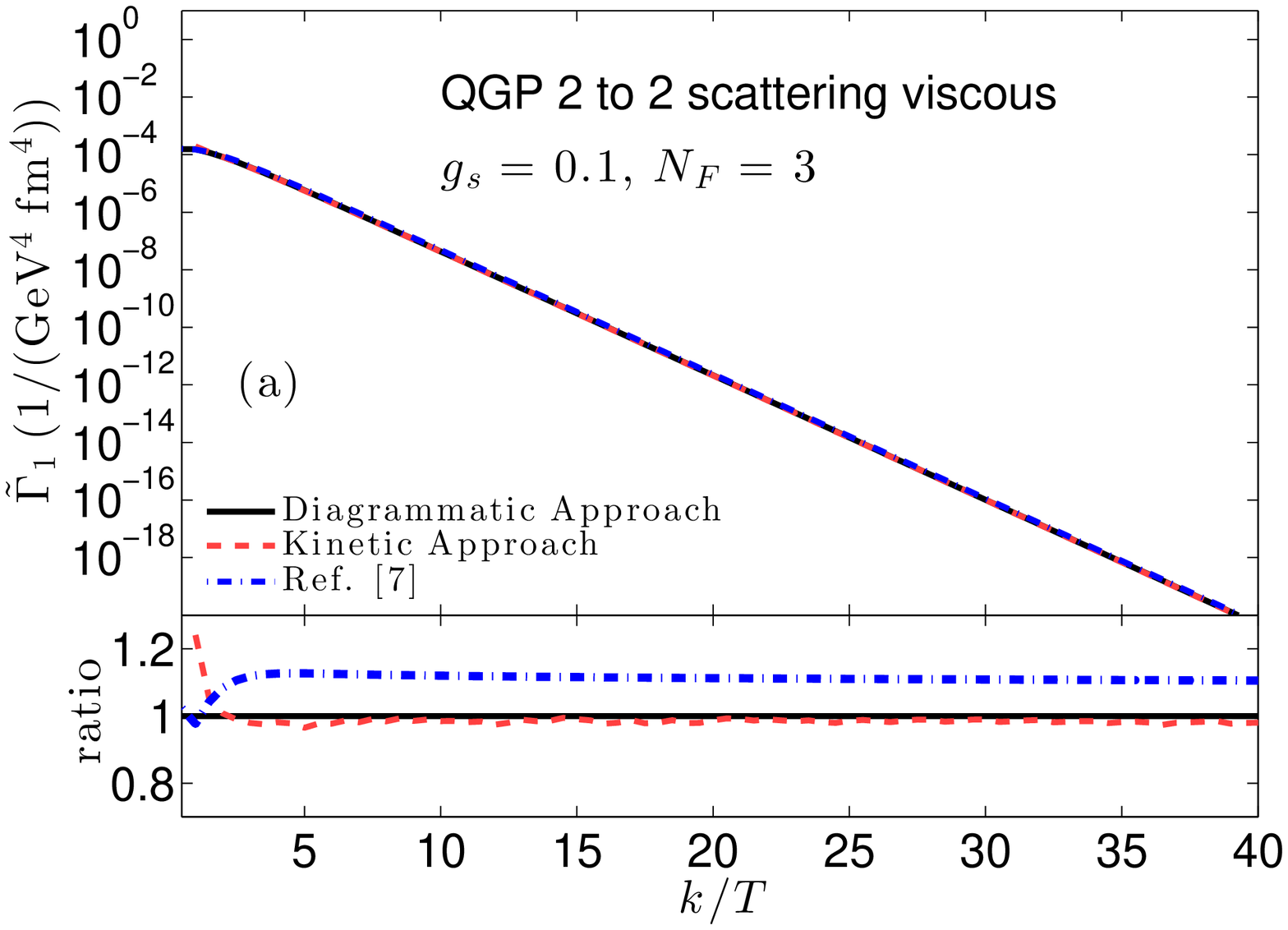} &
\includegraphics[width=0.48\linewidth]{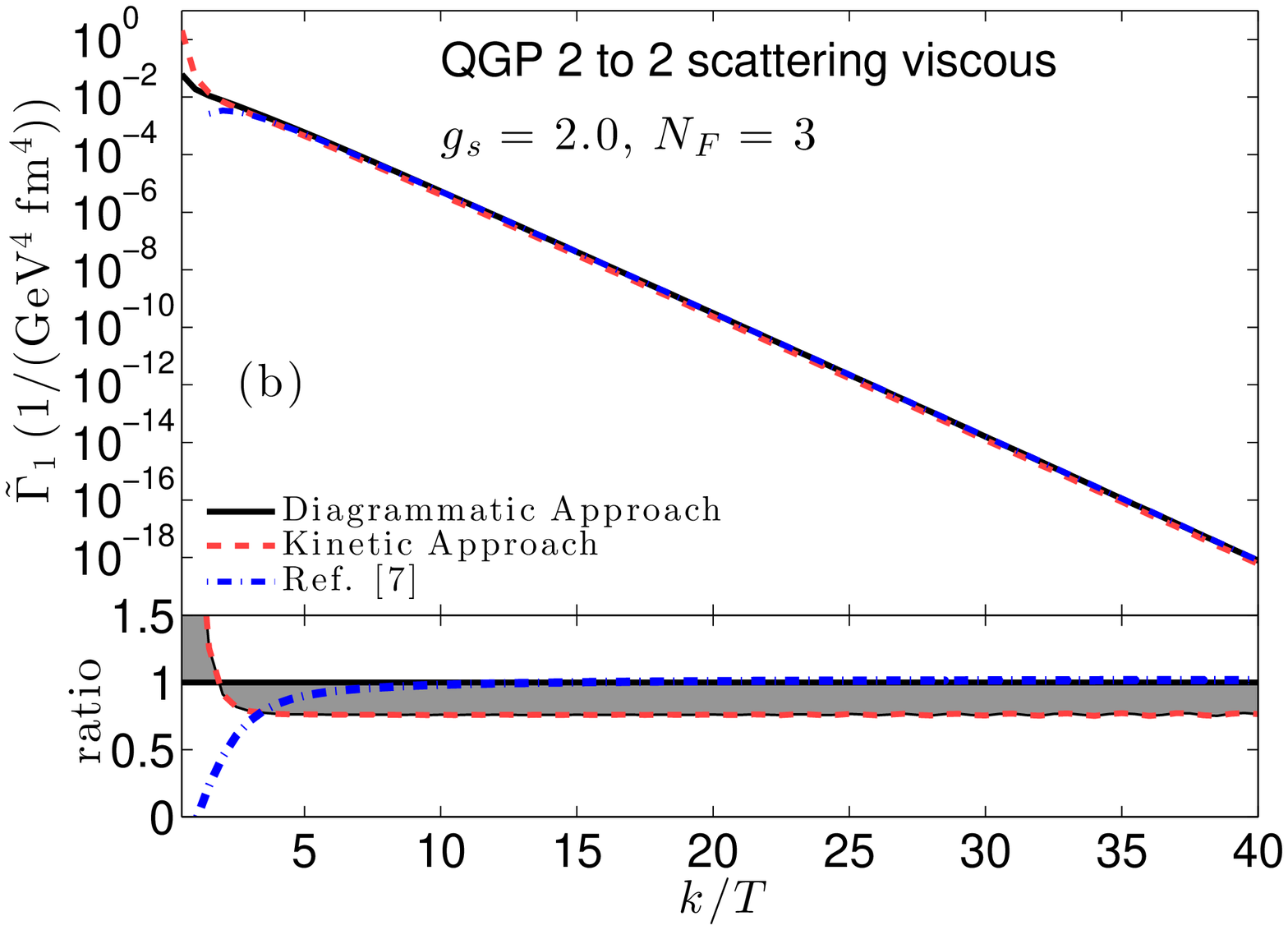}
\end{tabular}
\caption{(Color online) Similar to Fig.~\ref{fig4} but for the viscous correction coefficient $\tilde\Gamma_1$.}
\label{fig5}
\end{figure}
%================================================================================

Turning to the viscous correction coefficient $\tilde\Gamma_1$ shown in Fig.~\ref{fig5}, we see in panel (a) that at weak coupling ($g_s{\,=\,0.1}$) the result from the diagrammatic approach agrees well (within 2\%) with the kinetic approach for $k/T > 5$; significant deviations occur only when $k/T{\,<\,}2$. Ref.~\cite{Dusling:2009bc} again reproduces the correct $k/T$ dependence of $\tilde\Gamma_1$ but overestimates its absolute value by $\sim 10\%$, almost independent of $k/T$, compared to the other two approaches. We verified that this difference does not vanish in the limit $g_s{\,\to\,}0$, and must thus stem from either the forward scattering dominance approximation or the beyond-leading-log terms that were dropped in Ref.~\cite{Dusling:2009bc}. 

For larger coupling $g_s{\,=\,}2.0$ one observes a large degree of similarity between the $k/T$ dependences of the equilibrium rate $\tilde\Gamma_0$ (Fig.~\ref{fig4}b) and of the viscous correction coefficient $\tilde\Gamma_1$ (Fig.~\ref{fig5}b) as well as between the mutual relations among the different methods and approximations. The normalized viscous rate $\tilde\Gamma_1$ from the diagrammatic approach is systematically about 25\% larger than from the kinetic approach. For $k/T{\,\lesssim\,}1$ these two results begin to deviate significantly from each other. The rate from Ref.~\cite{Dusling:2009bc} yields about 25\% larger values for $\tilde\Gamma_1$ than the kinetic approach with full matrix elements. Its good agreement with the diagrammatic approach is however accidental, as discussed at the end of the previous section.
  
%%%%%%%%%%%%%%%%%%%%%%%%%%%%%%%%%%%%%%%%%%%%%
\subsection{Photon energy dependence of the ratio $\tilde\Gamma_1/\tilde\Gamma_0$}
\label{sec3c}
%%%%%%%%%%%%%%%%%%%%%%%%%%%%%%%%%%%%%%%%%%%%%

All results presented up to this point assumed a quadratic momentum dependence of the scalar function $\chi\left(\frac{k}{T}\right)$ parameterizing the deviation from local equilibrium in Eq.~(\ref{eq1}). Depending on the energy dependence of the scattering cross section between the medium constituents, the power 
%
%=============================== Fig. 8 =============================================
\begin{figure}[h]
\centering
\begin{tabular}{cc}
\includegraphics[width=0.435\linewidth]{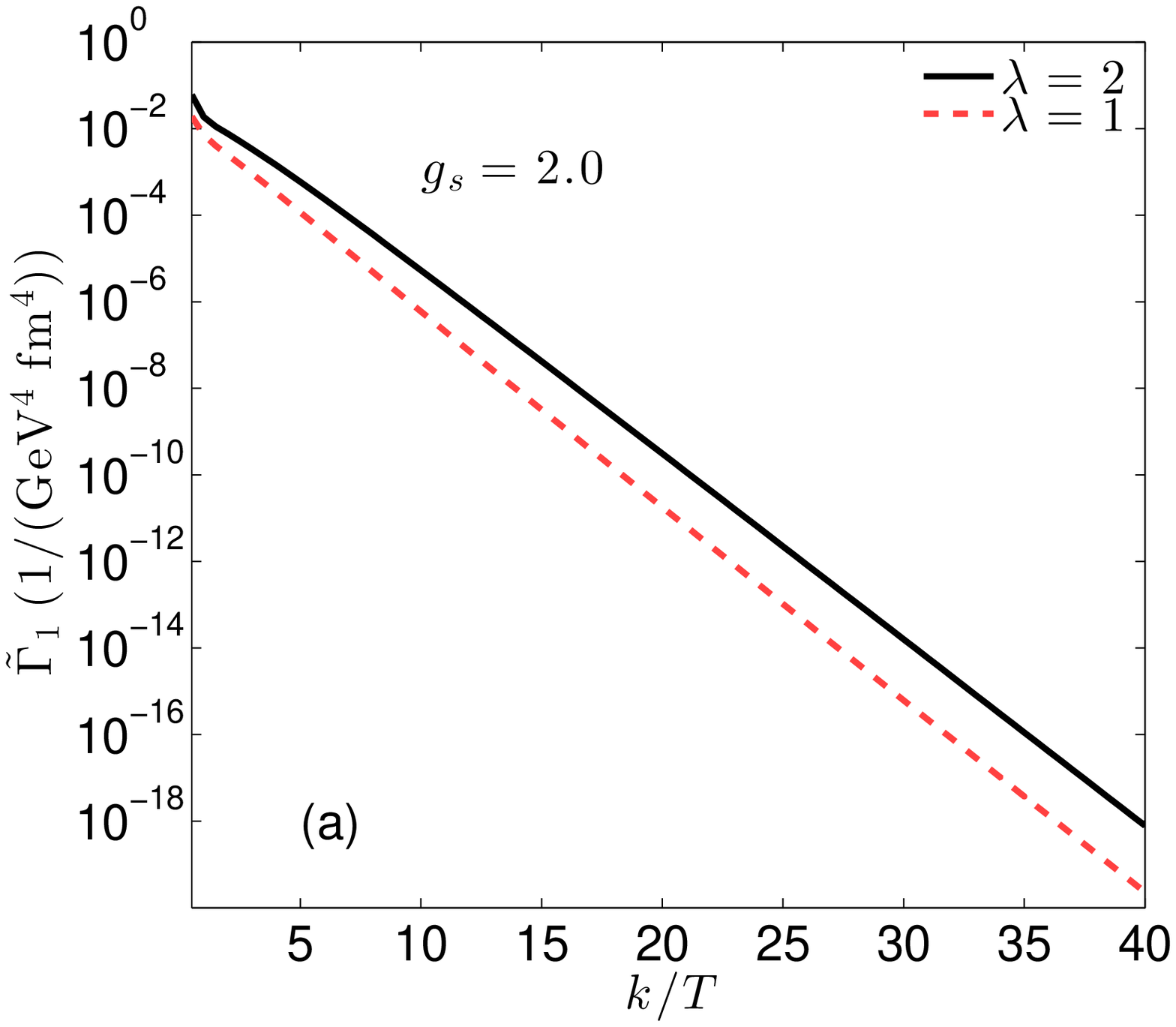} &
\includegraphics[width=0.475\linewidth]{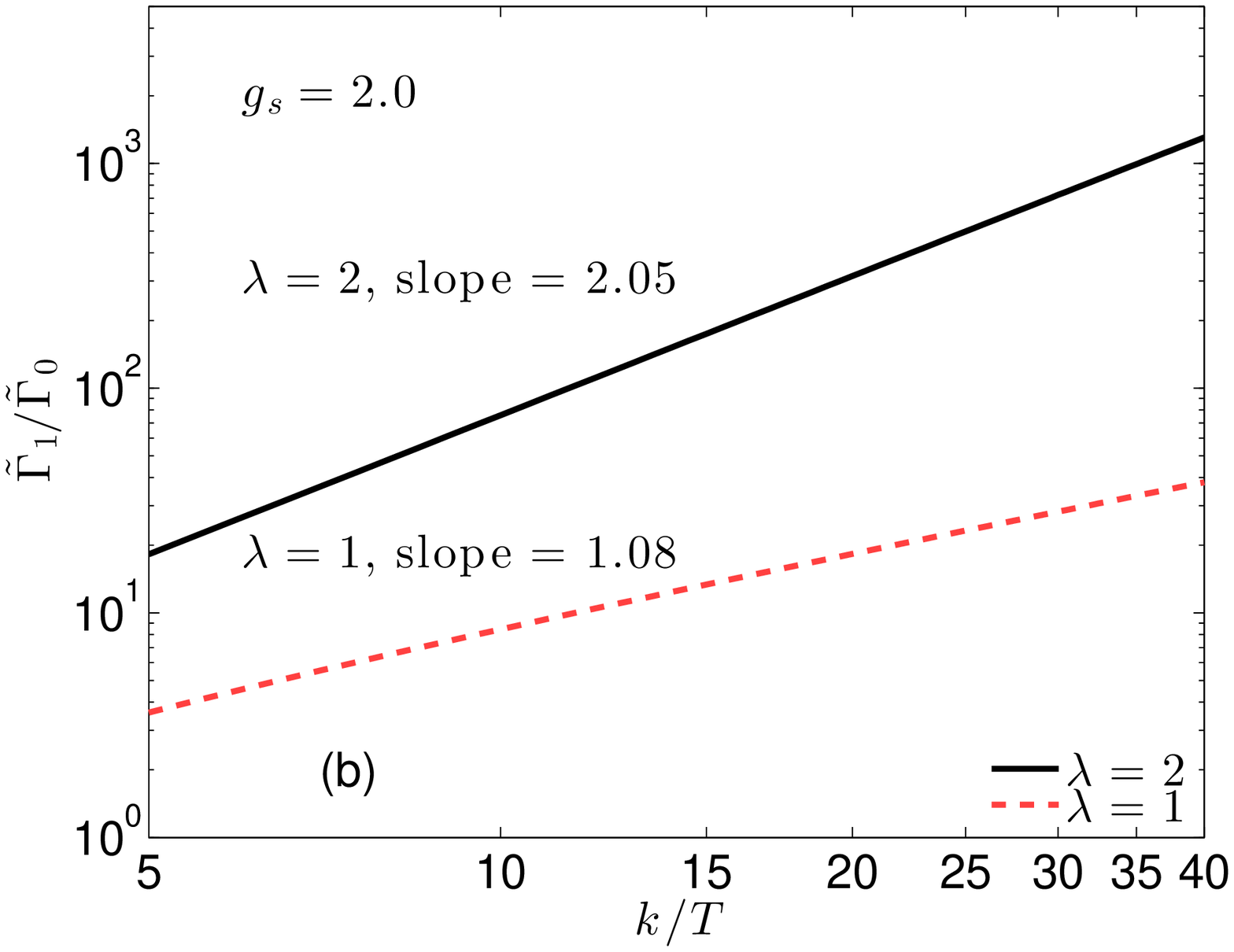}
\end{tabular}
\caption{(Color online) The viscous correction coefficient $\tilde\Gamma_1$ (a) and its ratio $\tilde\Gamma_1/\tilde\Gamma_0$ to the thermal equilibrium emission rate (b) as functions of $k/T$ at $g_s{\,=\,} 2.0$ for two values $\lambda$, $\lambda{\,=\,}1$ and 2 (see text for details). The slope parameters in panel (b) were obtained by a linear fit of the log-log plot for $k/T{\,>\,}5$.
}
\label{fig7}
\end{figure}
%==================================================================================
%
$\lambda$ of this momentum dependence typically spans the range between linear and quadratic \cite{Dusling:2009df}. In this section we therefore compare the viscous correction coefficient $\tilde\Gamma_1$ to the thermal equilibrium rate $\tilde\Gamma_0$ for $\lambda{\,=\,}1$ and $\lambda{\,=\,}2$, and explore the relationship between this power and the photon energy dependence of the ratio $\tilde\Gamma_1/\tilde\Gamma_0$ which parametrizes the relative importance of the viscous corrections to the thermal photon emission rate.

Figure~\ref{fig7} compares the viscous correction factors $\tilde\Gamma_1(k/T)$ obtained with $\chi(p/T){\,=\,}p/T$ and $\chi(p/T){\,=\,}(p/T)^2$. The thermal equilibrium rate in Fig.~\ref{fig4}b and the two different results for the viscous correction coefficient $\Gamma_1$ in Fig.~\ref{fig7}a for $k/T{\,>\,}5$ all fall roughly exponentially as functions of photon energy $k/T$, with very similar slopes. However, their ratio $\tilde\Gamma_1/\tilde\Gamma_0$ is revealed in Fig.~\ref{fig7}b to be a simple power $(k/T)^{\lambda'}$ of the scaled photon energy, with a power $\lambda'$ that reflects surprisingly closely the power $\lambda$ characterizing the energy dependence of the deviations $\delta f$. This power dependence is not trivial, but is in line with the result from Ref.~\cite{Dusling:2009bc}, which predicts the simple ratio $\tilde\Gamma_1/\tilde\Gamma_0=\left(1+f_{B 0}(k)\right) \chi(k/T)$.

%%%%%%%%%%%%%%%%%%%%%%%%%%%%%%%%%%%%%%%%%%%%%
\section{Conclusions}
\label{sec4}
%%%%%%%%%%%%%%%%%%%%%%%%%%%%%%%%%%%%%%%%%%%%%

In this work, we computed the photon emission rate from a quark-gluon plasma with locally anisotropic particle momentum distributions induced by a non-vanishing shear stress tensor. We calculated photon production from $2\rightarrow 2$ scattering processes in the QGP, with off-equilibrium corrections included to leading order in the shear stress. We employed both the diagrammatic and kinetic approaches to calculating the photon production rate and showed that the results agree in the weak coupling limit $g_s{\,\to\,}0$. The Feynman diagram based viscous rate calculation for processes involving soft scattering was considerably simplified by a proof of the KMS theorem for the exchanged quark propagator which was shown to hold, in the hard loop limit relevant for soft collisions, not only for the general type of shear-viscously deformed local momentum distributions, but for any local momentum distribution that is mirror symmetric under momentum reflection in the local rest frame.

We compared our equilibrium rates and viscous correction coefficients from both of these approaches with other existing results, specifically with the AMY parametrization of the thermal equilibrium photon emission rate \cite{Arnold:2001ms} and with the equilibrium rate and viscous correction factor obtained in Ref.~\cite{Dusling:2009bc} using simplified Compton and pair annihilation matrix elements evaluated in the forward scattering dominance approximation. In the diagrammatic approach we investigated the cutoff dependence of both the equilibrium rate and the viscous correction factor for both weak and moderately strong coupling. We found that both cutoff dependences are minimized by setting the cutoff to $q_\mathrm{cut}/T{\,\approx\,}\sqrt{g_s}$, but that for small $g_s{\,\ll\,1}$ there exists a wide ``window of insensitivity'' covering the range $g_sT{\,\ll\,}q_\mathrm{cut}/T{\,\ll\,}T$ where both the equilibrium emission rate $\tilde\Gamma_0$ and the viscous correction coefficient $\tilde\Gamma_1$ are approximately cutoff independent. Finally, we found that the photon energy dependence of the relative viscous correction $\tilde\Gamma_1/\tilde\Gamma_0$ to the photon emission rate is very close to the energy dependence of the off-equilibrium correction to the underlying quark and gluon distribution functions. 

The analysis presented in this work is restricted to $2\rightarrow 2$ collisions. A complete calculation that includes all contributions to the (viscously corrected) thermal photon emission rate at leading order in the strong coupling constant $g_s$ requires the inclusion and resummation of Bremsstrahlung processes induced by soft collinear collisions and of the Landau-Migdal-Pomeranchuk (LPM) interference effect \cite{Arnold:2001ms}. In thermal equilibrium these additional channels are known to boost the photon emission rate by about a factor of two over the $2\rightarrow 2$ collision processes discussed here. Work on computing the viscous corrections to these soft collinear photon production channels is ongoing and will be reported elsewhere.

%%%%%%%%%%%%%%%%%%%%%%%%%%%%%%%%%%%%%%%%%%%%%

\acknowledgments{The authors gratefully acknowledge clarifying discussions with Jacopo Ghiglieri, Aleksi Kurkela, Guy Moore, Bjoern Schenke and Mike Strickland. J.-F.P. thanks Gabriel Denicol for his help and critical comments. This work was supported in part by the U.S. Department of Energy under Grants No. DE-SC0004286 and (within the framework of the JET Collaboration) DE-SC0004104, and in part by the Natural Sciences and Engineering Research Council of Canada. J.-F.P. acknowledges support through grants from Hydro-Qu\'{e}bec and the Fonds de Recherche du Qu\'{e}bec}, and C. G. acknowledges support from the Hessian initiative for excellence (LOEWE) through the Helmholtz International Center for FAIR (HIC for FAIR).

%%%%%%%%%%%%%%%%%%%%%%%%%%%%%%%%%%%%%%%%%%%%%

%%%%%%%%%%%%%%%%%%%%%%%%%%%%%%%%%%%%%%%%%%%%%
\appendix
%%%%%%%%%%%%%%%%%%%%%%%%%%%%%%%%%%%%%%%%%%%%%
\section{Photon production by large angle scattering}
\label{appHard}
%%%%%%%%%%%%%%%%%%%%%%%%%%%%%%%%%%%%%%%%%%%%%

We first treat the $t$-channel terms. We define $\bm{q}{\,=\,}\bm{p}{-}\bm{k}$ and $\omega{\,=\,}p{-}k$ (where $p{\,=\,}|\bm{p}|$ etc.) such that an infrared cut-off can be placed on the exchanged momentum $q$. Using the energy-momentum $\delta$-function to eliminate phase-space integrals is easiest by first considering the momentum-integrated photon emission rate for a single scattering channel:
\begin{equation}
   R = \int_{p, p', k, k'} (2\pi)^4 \delta^{(4)}(P{+}P'{-}K{-}K') \vert \mathcal{M} \vert^2 f(P) f(P') (1 \pm f(K')).
\label{hard.2}
\end{equation}
The integrals are most easily evaluated in the local fluid rest frame, using a coordinate system with its $z$ axis aligned with the photon momentum $\bm{k}$ and the $x{-}z$ plane spanned by $\bm{k}$ and $\bm{q}$. In this frame, the integrand in Eq.~(\ref{hard.2}) is determined by the momentum magnitudes $p$, $p'$, and $k$ and three angles, $\theta_{kq}$, $\theta_{p'q}$, and $\phi_{p'}$. The remaining angular integrals give trivial factors. We use $\delta^{(3)}(\bm{p}{+}\bm{p'}{-}\bm{k}{-}\bm{k'})$ to perform the integration over $\bm{k'}$:
\begin{eqnarray}
   R &=& \int p'^2 dp' q^2 dq\, k^2 dk\, d\cos\theta_{kp}\, d\cos\theta_{p'q}\, d\phi_{p'} 
   \frac{2(2\pi)^2}{(2\pi)^8 2^4 p p' k k'}\, \delta(p{+}p'{-}k{-}k').
\notag \\ 
   && \times\, \vert \mathcal{M} \vert^2 f(p) f(p') \bigl(1{\pm}f(p{+}p'{-}k)\bigl)
\label{hard.3}
\end{eqnarray}
The remaining $\delta$-function is split in two by introducing a dummy integration:
\begin{equation}
   \delta(p{+}p'{-}k{-}k') = \int_{-\infty}^{+\infty} d \omega\, \delta( \omega{+}k{-}p)\, 
   \delta(\omega{+}p'{-}k'),
\end{equation}
with each factor rewritten to perform one of the polar angle integrations:
\begin{eqnarray}
   \delta(\omega{+}k{-}p') &=& \frac{p}{q k}\, 
   \delta\left(\cos\theta_{q k} - \frac{\omega^2{-}q^2{+}2 \omega k}{2qk}\right) \theta(\omega{+}k),
\label{hard.4a}
\\
   \delta(\omega{+}p'{-}k') &=& \frac{k'}{q p'}\, 
   \delta\left(\cos\theta_{p' q} - \frac{\omega^2{-}q^2{+}2 \omega p'}{2p'q}\right) \theta(\omega{+}p').
\label{hard.4b}
\end{eqnarray}
Doing so yields
\begin{equation}
   R = \int dq\, dp'\, dk\, d\omega\, d\phi_{p'}\, \frac{1}{8(2\pi)^6} \vert \mathcal{M} \vert^2 
   f(\omega{+}k) f(p') \bigl(1{\pm}f(\omega{+}p')\bigr) \theta(\omega{+}k) \theta(\omega{+}p').
\end{equation}
Now we can return to the differential photon emission rate for the selected channel:
\begin{equation}
   k\frac{dR}{d^3 k} = \frac{1}{16(2\pi)^7 k} \int dq\, dp'\, d\omega d\phi_{p'}  
   |\mathcal{M}|^2 f(\omega{+}k) f(p') \bigl(1{\pm}f(\omega{+}p')\bigr) \theta(\omega{+}k) 
   \theta(\omega{+}p').
\label{hard.5}
\end{equation}
The Mandelstam variables in the matrix elements are expressed in terms of these integration variables as
\begin{eqnarray}
  &&t = \omega^2 - q^2, \qquad s=-t-u,
\label{hard.6a}
\\
   &&u = - 2 p' k (1 - \cos\theta_{kq} \cos\theta_{p'q} + \sin\theta_{kq} \sin\theta_{p'q} \cos\phi_{p'}),
\label{hard.6b}
\end{eqnarray}
with $\cos\theta_{kq}$ and $\cos\theta_{p'q}$ given by the poles of the $\delta$-functions in Eqs.~(\ref{hard.4a}) and (\ref{hard.4b}). 

With our anisotropic distribution function, Eq. (\ref{eq1}), the integral over $\phi_{p'}$ can be done analytically. Splitting $f{\,=\,}f_0{+}\delta f$ and ignoring all $\delta f$ terms we obtain the equilibrium rate $\Gamma_0$ in Eq.~(\ref{eq2}) which, after adding all three $t$-channel contributions, summing over quark species $s$ and over quark- and antiquark contributions to the Compton channel, reads
\begin{eqnarray}
\label{A10}
   \Gamma_0 &=& \frac{{\cal N}}{16(2\pi)^6 k} 
   \int_{q_\mathrm{cut}}^{+\infty} dq 
   \int_{\mathrm{max}\{q - 2k, -q\}}^q d\omega \int_{(q - \omega)/2}^{+\infty} dp' 
\notag \\ 
   && \times \Bigl[ \Bigl(1 - \frac{2 p' k}{\omega^2{-}q^2} (1 - \cos\theta_{kq} \cos\theta_{p'q}) \Bigr) 
   f_{F0}(\omega{+}k) f_{B0}(p') \bigl(1{-}f_{F0}(p'{+}\omega)\bigr) 
\notag \\
   &&\quad\  - \frac{2 p' k}{\omega^2{-}q^2} (1 - \cos\theta_{kq} \cos\theta_{p'q}) f_{F0}(\omega{+}k) 
   f_{F0}(p') \bigl(1{+}f_{B0}(p'{+}\omega)\bigr) \Bigr],
\end{eqnarray}
where we implemented the infrared cutoff $q_\mathrm{cut}$ in the $q$ integral, and where
\begin{equation}
  {\cal N} = 16  N_C C_F e^2 g_s^2 \sum_s^{N_f} q_s^2\ 
  = 2^8 \pi N_C \alpha_\mathrm{EM} \frac{m_\infty^2}{T^2} \sum_s^{N_f} q_s^2.
\end{equation}
In (\ref{A10}) the first term in the square brackets accounts for Compton scattering, the second for $q\bar{q}$ annihilation. 

Now we add all contributions linear in $\delta f$, write the result as in Eq.~(\ref{2.A.4}) and read off the coefficient $\Gamma^{\mu\nu}$. Contracting with $a_{\mu\nu}$ to obtain the viscous correction coefficient $\Gamma_1{\,=\,}a_{\mu\nu}\Gamma^{\mu\nu}$ in Eq.~(\ref{eq2}) we get from the $-\frac{s}{t}$
part of $|{\cal M}|^2$ the Compton scattering contribution
\begin{eqnarray}
   \Gamma_1^{(-s/t)} &=& \frac{\cal N}{16(2\pi)^6 k} 
   \int_{q_\mathrm{cut}}^{+\infty} dq \int_{\mathrm{max}\{q - 2k, -q\}}^q d\omega 
   \int_{(q - \omega)/2}^{+\infty} dp'\,  f_{F0}(\omega{+}k)\, f_{B0}(p')\, \bigl(1{-}f_{F0}(p'{+}\omega)\bigr) 
\notag \\ 
   &\times& \Bigl\{\Bigl(1 - \frac{2 p' k}{\omega^2{-}q^2} (1 - \cos\theta_{kq} \cos\theta_{p'q}) \Bigr) 
\notag \\
   &&\quad \times\Bigl[\bigl(1{-}f_{F0}(\omega{+}k)\bigr) \chi\Bigl(\frac{\omega{+}k}{T}\Bigr) 
   \Bigl(-\frac{1}{2} + \frac{3}{2}\Bigl(\frac{q \cos\theta_{kq}{+}k}{\omega{+}k} \Bigr)^2\Bigr) 
\notag \\
   &&\qquad - f_{F0}(p'{+}\omega) \chi\Bigl(\frac{p'{+}\omega}{T}\Bigr) 
        \Bigl(-\frac{1}{2} + \frac{3}{2}\frac{1}{(p'{+}\omega)^2} 
        \bigl( (p' \cos\theta_{p'q} + q)^2\cos^2\theta_{kq} 
\notag \\        
    &&\hspace*{8.6cm}   
        +{\textstyle\frac{1}{2}}p'^2\sin^2\theta_{kq} \sin^2\theta_{p'q}\bigr)\Bigr)  
\notag \\
   &&\qquad + \bigl(1{+}f_{B0}(p')\bigr) \chi\Bigl(\frac{p'}{T}\Bigr) 
        \Bigl(-\frac{1}{2} + \frac{3}{2}\bigl(\cos^2\theta_{kq} \cos^2\theta_{p'q} + 
        {\textstyle\frac{1}{2}}\sin^2\theta_{kq} \sin^2\theta_{p'q}\bigr) \Bigr) \Bigr]
\notag \\
   &&\quad +\frac{2 p' k}{\omega^2{-}q^2} \sin\theta_{kq} \sin\theta_{p'q} 
\notag \\
   &&\quad \times \Bigl[ \bigl(1{+}f_{B0}(p')\bigr) \chi\Bigl(\frac{p'}{T}\Bigr)  
        \frac{3}{2} \cos\theta_{kq} \cos\theta_{p'q} \sin\theta_{kq} \sin\theta_{p'q} \notag \\
&&\quad-f_{F0}(p'{+}\omega) \chi\Bigl(\frac{p'{+}\omega}{T}\Bigr) \frac{3}{2} \frac{1}{(p'{+}\omega)^2} 
    \bigl(p' \sin\theta_{kq} \sin\theta_{p'q}\cos\theta_{kq}(p' \cos\theta_{p'q}{+}q)\bigr) \Bigr] \Bigr\},\qquad
\end{eqnarray}
while the $\frac{u}{t}$ part gives the contribution from $q\bar{q}$ annihilation: 
\begin{eqnarray}
   \Gamma_1^{(u/t)} &=& \frac{\cal N}{16(2\pi)^6 k} \int_{q_\mathrm{cut}}^{+\infty} dq 
   \int_{\mathrm{max}\{q - 2k, -q\}}^q d\omega \int_{(q - \omega)/2}^{+\infty} dp'\,  
   f_{F0}(\omega{+}k)\, f_{F0}(p')\, \bigl(1{+}f_{B0}(p'{+}\omega)\bigr)
\notag \\ 
   &\times& \Bigl(- \frac{2 p' k}{\omega^2{-}q^2} \Bigr) \Bigl\{( 1 - \cos\theta_{kq} \cos\theta_{p'q}) 
\notag \\
   &\times&\Bigl[\bigl(1{-}f_{F0}(\omega{+}k)\bigr) \chi\Bigl(\frac{\omega{+}k}{T}\Bigr) 
   \Bigl(-\frac{1}{2} + \frac{3}{2}\Bigl(\frac{q \cos\theta_{kq}{+}k}{\omega{+}k} \Bigr)^2\Bigr)
\notag \\
   && +\bigl(1{-}f_{F0}(p')\bigr) \chi\Bigl(\frac{p'}{T}\Bigr) 
        \Bigl(-\frac{1}{2} 
        + \frac{3}{2}\bigl(\cos^2\theta_{kq} \cos^2\theta_{p'q} + {\textstyle\frac{1}{2}}\sin^2\theta_{kq}
           \sin^2\theta_{p'q} \bigr)\Bigr) 
\notag \\
     && + f_{B0}(p'{+}\omega) \chi\Bigl(\frac{p'{+}\omega}{T}\Bigr) 
     \Bigl(-\frac{1}{2} + \frac{3}{2}\frac{1}{(p'{+}\omega)^2} 
     \bigl((p' \cos\theta_{p'q} + q)^2\cos^2\theta_{kq} 
\notag \\
     &&\hspace*{8cm}     
     + {\textstyle\frac{1}{2}}p'^2\sin^2\theta_{kq} \sin^2\theta_{p'q}\bigr)\Bigr) \Bigr] 
\notag \\
   &+& \sin\theta_{kq} \sin\theta_{p'q} 
\notag \\
   &\times& \Bigl[\bigl(1{-}f_{F0}(p')) \chi\Bigl(\frac{p'}{T}\Bigr) 
                  \frac{3}{2} \cos\theta_{kq} \cos\theta_{p'q} \sin\theta_{kq} \sin\theta_{p'q} 
\notag \\
   &+& f_{B0}(p'{+}\omega) \chi\Bigl(\frac{p'{+}\omega}{T}\Bigr) 
   \frac{3}{2} \frac{1}{(p'{+}\omega)^2} \bigl(p' \sin\theta_{kq} \sin\theta_{p'q}\cos\theta_{kq}
   (p' \cos\theta_{p'q}{+}q)\bigr) \Bigr] \Bigr\}.\quad
\end{eqnarray}
For the $s$-channel diagrams we define $\bm{q}{\,=\,}\bm{p}{+}\bm{p'}$ and $\omega{\,=\,}p{+}p'$ and follow the same procedure:
\begin{eqnarray}
   \Gamma_0 &=& \frac{\cal N}{16(2\pi)^6 k} \int_{k}^{+\infty} d\omega 
   \int_{\vert 2k - \omega \vert}^\omega dq \int_{(\omega - q)/2}^{(\omega + q)/2} dp' 
   \notag \\ 
  &&\times\, \frac{2 p' k}{\omega^2{-}q^2}\, (1{-}\cos\theta_{kq} \cos\theta_{p'q})\, 
  f_{B0}(\omega{-}p')\, f_{F0}(p')\, \bigl(1{-}f_{F0}(\omega{-}k)\bigr),
\end{eqnarray}
where now $\cos \theta_{kq}{\,=\,}\frac{q^2 - \omega^2 + 2 \omega k}{2qk}$ and $\cos \theta_{p'q}{\,=\,}\frac{q^2 - \omega^2 + 2 \omega p'}{2qp'}$. The $s$-channel contribution to the viscous correction coefficient is
\begin{eqnarray}
  \Gamma_1 &=& \frac{\cal N}{16(2\pi)^6 k} \int_{k}^{+\infty} d\omega 
  \int_{\vert 2k - \omega \vert}^\omega dq \int_{(\omega - q)/2}^{(\omega + q)/2} dp'\, 
  f_{B0}(\omega{-}p')\, f_{F0}(p') \bigl(1{+}f_{F0}(\omega{-}k)) 
\notag \\ 
   &\times& \frac{2 p' k}{\omega^2{-}q^2} \Bigl\{ (1{-}\cos\theta_{kq} \cos\theta_{p'q})
\notag \\
  &\times&\Bigl[\bigl(1{-}f_{F0}(p')\bigr) \chi\Bigl(\frac{p'}{T}\Bigr) 
  \Bigl(-\frac{1}{2} + \frac{3}{2}\bigl(\cos^2\theta_{kq} \cos^2\theta_{p'q} 
  + {\textstyle\frac{1}{2}}\sin^2\theta_{kq} \sin^2\theta_{p'q} \bigr)\Bigr)   
\notag \\
  && - f_{F0}(\omega{-}k)\chi\Bigl(\frac{\omega{-}k}{T}\Bigr) 
   \Bigl(-\frac{1}{2} + \frac{3}{2} \Bigr(\frac{q \cos\theta_{kq}{-}k}{\omega{-}k} \Bigr)^2\Bigr)
\notag \\
   && + \bigl(1{+}f_{B0}(\omega{-}p')\bigr) \chi\Bigl(\frac{\omega{-}p'}{T}\Bigr)
    \Bigl(-\frac{1}{2} + \frac{3}{2}\frac{(q{-}p' \cos\theta_{p'q})^2\cos^2\theta_{kq} 
                                                         + \frac{1}{2}p'^2\sin^2\theta_{kq} \sin^2\theta_{p'q}}{(\omega - p')^2}
                                                          \Bigr)  \Bigr]
\notag \\
   &+& \sin\theta_{kq} \sin\theta_{p'q} 
\notag \\
   &\times& \Bigl[ - \bigl(1{-}f_{F0}(p')\bigr)\, \chi\Bigl(\frac{p'}{T}\Bigr) \, 
     \frac{3}{2} \cos\theta_{kq} \cos\theta_{p'q} \sin\theta_{kq} \sin\theta_{p'q} 
\notag \\
   &&+ \bigl(1{+}f_{B0}(\omega{-}p')\bigr)\,\chi\Bigl(\frac{\omega{-}p'}{T}\Bigr)\, 
   \frac{3}{2} \frac{p' \sin\theta_{kq} \sin\theta_{p'q}\cos\theta_{kq}(q{-}p' \cos\theta_{p'q})}
                           {(\omega{-}p')^2}  \Bigr] \Bigr\}.\quad
\end{eqnarray}
The remaining three integrals are straightforward to evaluate numerically, using e.g. Gaussian quadrature.

\section{Parameterization of the ideal and viscous photon rates}
\label{appendixFit}

We wrote our final result for the photon emission rate as
\begin{equation*}
k \frac{dR}{d^3k} = T^2 \left( \tilde\Gamma_0 + 
                                               \frac{\pi^{\mu\nu} \hat{k}_\mu \hat{k}_\nu}{2(e{+}\p)}\,
                                               \tilde\Gamma_1\right)
\end{equation*}

We described in Sec.~\ref{sec2} how to evaluate $\tilde\Gamma_0$ and $\tilde\Gamma_1$. We provide here a parametrization of these two functions as computed in the diagrammatic approach, described in Sec.~\ref{sec2b}. We used the prescription $q_{\textrm cut}/T=\sqrt{g_s}$ to fix the cut-off, since $\sqrt{g_s}$ is in general close to the minimum of the cut-off dependence. The parametrization was made with $\chi(k/T) = (k/T)^2$ ($\lambda=2$) in (\ref{eq1}), corresponding to a quadratic dependence in the energy of the momentum anisotropy ansatz.

We write both $\tilde\Gamma_0$ and $\tilde\Gamma_1$ as
\begin{eqnarray}
  \tilde\Gamma_0(k/T) & = & \mathcal{B}(k/T) \exp \left\{ F\left(\ln(g_s),\ln(k/T)\right) \right\} \nonumber \\
	\tilde\Gamma_1(k/T) & = & \mathcal{B}(k/T) \exp \left\{ G\left(\ln(g_s),\ln(k/T) \right) \right\}
\end{eqnarray}
with
\begin{equation}
\mathcal{B}(k/T)=\frac{2 \alpha_{E M} g_s^2}{(2 \pi)^3}  \left[ \sum_s^{N_f} q_s^2 \right]  n_f(k/T)
\end{equation}
and $F$ an $G$ being given by the following parameterizations:
\begin{equation}
\begin{array}{llllllllllllll}
F(x,y) & = &  & [ 0.200 & - & 0.607& x & - & 0.131& x^2 & + & 0.0242& x^3 ] &  \\
	& & + & [ 0.0574 &+& 0.0359& x &-& 0.219& x^2 &-& 0.00280& x^3 ] & y  \\
	& & + & [ 0.0978 &+& 0.00833& x &+& 0.445& x^2 &-& 0.106& x^3 ] & y^2  \\
	& & + & [ 0.0167 &+& 0.0389& x &-& 0.294& x^2 &+& 0.112& x^3 ] & y^3  \\
	& & + & [ -0.018749 &-& 0.0196& x &+& 0.0837& x^2 &-& 0.0396& x^3 ] & y^4   \\
	& & + & [ 0.00279 &+& 0.00257& x &-& 0.00873& x^2 &+& 0.00464& x^3 ] & y^5 
\end{array}
\end{equation}

\begin{equation}
\begin{array}{llllllllllllll}
G(x,y) & = &  & [ -0.314 &-& 0.967&x&+& 0.0159 &x^2&+& 0.250    &x^3 ] &  \\
     	& & + & [ 2.53    &+& 1.08  &x&-& 2.07   &x^2&+& 0.195 &x^3 ] & y  \\
     	& & + & [ -0.299  &-& 1.41  &x&+& 3.73   &x^2&-& 1.10  &x^3 ] & y^2  \\
     	& & + & [ 0.213   &+& 0.879 &x&-& 2.52   &x^2&+& 0.946 &x^3 ] & y^3  \\
     	& & + & [ -0.0678 &-& 0.244 &x&+& 0.733  &x^2&-& 0.307 &x^3 ] & y^4   \\
     	& & + & [ 0.00758 &+& 0.0247&x&-& 0.077  &x^2&+& 0.0343&x^3 ] & y^5 
\end{array}
\end{equation}

The parameterizations are accurate to within $3\%$ in the range $1<g_s<3.1$ and \mbox{$1<k/T<30$}.

%idealRateFactor[kOverT_, gs_] := 
 %Exp[0.200024 - 0.606887 Log[gs] - 0.131303 Log[gs]^2 + 
   %0.0242216 Log[gs]^3 + (0.0574223 + 0.0358792 Log[gs] - 
      %0.219267 Log[gs]^2 - 0.00280486 Log[gs]^3) Log[
     %kOverT] + (0.0978125 + 0.00833357 Log[gs] + 0.445448 Log[gs]^2 - 
      %0.105617 Log[gs]^3) Log[kOverT]^2 + (0.0166855 + 
      %0.038851 Log[gs] - 0.29352 Log[gs]^2 + 0.111947 Log[gs]^3) Log[
      %kOverT]^3 + (-0.018749 - 0.0195898 Log[gs] + 
      %0.0837425 Log[gs]^2 - 0.0395994 Log[gs]^3) Log[
      %kOverT]^4 + (0.00278864 + 0.00257356 Log[gs] - 
      %0.00872805 Log[gs]^2 + 0.00464229 Log[gs]^3) Log[kOverT]^5]
%
%viscousRateFactor[kOverT_, gs_] := 
 %Exp[-0.314322 - 0.967462 Log[gs] + 0.0159076 Log[gs]^2 + 
   %0.249609 Log[gs]^3 + (2.52998 + 1.08161 Log[gs] - 
      %2.07468 Log[gs]^2 + 0.194961 Log[gs]^3) Log[
     %kOverT] + (-0.298736 - 1.40953 Log[gs] + 3.72824 Log[gs]^2 - 
      %1.09869 Log[gs]^3) Log[kOverT]^2 + (0.212533 + 
      %0.879362 Log[gs] - 2.5181 Log[gs]^2 + 0.946294 Log[gs]^3) Log[
      %kOverT]^3 + (-0.0677817 - 0.243845 Log[gs] + 
      %0.733307 Log[gs]^2 - 0.306536 Log[gs]^3) Log[
      %kOverT]^4 + (0.00757655 + 0.0247427 Log[gs] - 
      %0.077341 Log[gs]^2 + 0.0342706 Log[gs]^3) Log[kOverT]^5]

%%%%%%%%%%%%%%%% References %%%%%%%%%%%%%%%%%%%%%%%%%%%
%

%\bibliographystyle{h-physrev3}
\bibliographystyle{ieeetr}
\bibliography{biblio}

\end{document}